\documentclass[10pt]{iopart}
\usepackage{amssymb}

\newcommand{\re}{\mathop{\mathrm{Re}}}
\newcommand{\im}{\mathop{\mathrm{Im}}}
\newcommand{\diag}{\mathop{\mathrm{diag}}}

\begin{document}

\title[Random matrices close to Hermitian or unitary]{RANDOM MATRICES CLOSE TO HERMITIAN OR UNITARY:\\ overview of methods and results }

\author{Yan V. Fyodorov$^{\S}$ and H.-J.Sommers$^{\P}$}

\address{$\S$ Department of Mathematical Sciences,
Brunel University, Uxbridge UB83PH, United Kingdom}

\address{
$\P$ Fachbereich Physik, Universit\"at  Essen,
D-45117 Essen, Germany}

\date{published in: \\
{\bf J.Phys.A} v.36 (2003), 3303-3348 ;
{\sf Special Issue on Random Matrix Theory }, 2003\\
Ed. by P. Forrester, N. Snaith and J.J.M. Verbaarschot}
%\maketitle

\begin{abstract}
The paper discusses recent progress in understanding
 statistical properties of eigenvalues of (weakly)
non-Hermitian and non-unitary random matrices. The first type of ensembles is of the form $\hat{J}=\hat{H}-i\hat{\Gamma}$, with $\hat H$ being
 a large random $N\times N$ {\it Hermitian} matrix with
independent entries "deformed" by a certain {\it anti-Hermitian}
$N\times N$ matrix $i\hat \Gamma$ satisfying in the limit of large dimension
$N$
the condition: $\mbox{Tr} {\hat H}^2 \propto N  \mbox{Tr} {\hat \Gamma}^2$.
Here $\hat \Gamma$ can be either a random or just a fixed given
Hermitian matrix. Ensembles of such a type with $\hat \Gamma\ge 0$ emerge naturally when
describing quantum scattering in systems with chaotic dynamics
and serve to describe resonance statistics. Related
models are used to mimic complex spectra of the Dirac operator with
chemical potential in the context of Quantum Chromodynamics.

Ensembles of the second type, arising naturally in scattering
theory of {\it discrete-time} systems,
are formed by $N\times N$ matrices $\hat{A}$
with complex entries such that
$\hat{A}^{\dagger}\hat{A}=\hat{I}-\hat{T}$. For $\hat{T}=0$
this coincides with the Circular Unitary Ensemble, and $0 \le\hat{T}\le \hat{I}$
describes deviation from unitarity. Our result amounts to
answering statistically the following old question:
given the singular values of a matrix $\hat{A}$
 describe the locus of its eigenvalues.

We systematically show that the obtained expressions for the
correlation functions of complex eigenvalues describe a non-trivial
crossover from Wigner-Dyson statistics of real/unimodular
eigenvalues
typical for Hermitian/unitary matrices to   Ginibre statistics in the
complex plane typical for ensembles with strong non-Hermiticity:
$<\mbox{Tr} {\hat H}^2>\propto< \mbox{Tr} {\hat \Gamma}^2>$ when $N\to \infty$. Finally we discuss (scarce) results available on
eigenvector statistics for weakly non-Hermitian random matrices.
\end{abstract}

\section{Introduction}

As is well-known, the statistics of highly excited
bound states of {\it closed} quantum chaotic systems
of quite different microscopic nature is universal. Namely, it
turns out to be independent of
the microscopic details when sampled on   energy intervals large in
comparison with the mean level spacing $\Delta$, but smaller
than the so called Thouless energy scale.
The latter is related by the Heisenberg uncertainty principle to the
relaxation time necessary for the classically chaotic system to reach
equilibrium in  phase space \cite{RMT1}.
Moreover, the
spectral correlation functions turn out to
be exactly those which are provided by the theory of
large random Hermitian matrices with independent, identically
distributed Gaussian entries. The correspondence holds in the limit
of large matrix
 dimension
 on the so-called {\it local} scale.
The local scale is determined by the typical
separation $\Delta=\langle X_i-X_{i-1}\rangle$
between neighbouring eigenvalues situated
around a point $X$, with the brackets standing for the statistical
averaging.
 Microscopic arguments supporting the use of random matrices
for describing the universal properties of quantum chaotic systems have
been
provided recently by several groups, based both on   traditional
semiclassical
periodic orbit expansions \cite{per,bogomol} and on advanced
field-theoretical methods \cite{MK,aaas}.

In parallel, it was realized that another type of random
matrices - those with the so-called chiral structure - serve as a
pattern for
universal statistics of low-lying eigenvalues of the
Euclidean Dirac operator.
The latter is of crucial importance for describing such
a fundamental
phenomenon as spontaneous breakdown of chiral symmetry in Quantum
Chromodynamics (QCD). That line of development is thoroughly
described in \cite{VW}.
The corresponding (anti-)Hermitian
matrices are of the form
\begin{equation}\label{dirac}
\hat{D}=i\left(\begin{array}{cc}{\bf 0}&\hat{J}\\ \hat{J}^{\dagger}&
{\bf 0} \end{array}\right)
\end{equation}
where $\hat J$ stands for a complex matrix, with $\hat{J}^{\dagger}$ being
its Hermitian conjugate. Such an off-diagonal block structure is
characteristic for systems with chiral symmetry.
One of the main objects of interest in QCD is the so-called partition
function  used to describe a system of quarks characterized by
$N_F$ flavours and quark
masses $m_f$ interacting with the Yang-Mills gauge fields.
At the level of Random Matrix Theory the true partition function
is replaced by the matrix integral:
\begin{equation}\label{partf}
{\cal Z}_{N_F}(m)=\int d\hat{J}d\hat{J}^{\dagger}\ \prod_{f=1}^{N_F}\det\{
\hat{D}+m_f\}\
e^{- N \mbox{\small Tr}\hat{J}^{\dagger}\hat{J}}
\end{equation}
Here instead of an integration over gauge fields in the
QCD partition function we have the integration over matrices
$\hat{J}$ \cite{VW}. It represents, in fact, the
ensemble average of the product of characteristic polynomials of
$\hat{D}$.

The Hermitian ensemble of block off-diagonal matrices $-i\hat{D}$
 with independent,
identically distributed Gaussian entries is
known as the {\it chiral} Gaussian unitary ensemble (chGUE).
The eigenvalues of chiral
matrices appear in pairs $\pm \lambda _k\,\,,\,\, k=1,...,N$.
It is evident that the nonzero $\epsilon_k=\lambda_k^2\ne 0$
are the singular values of the matrix $\hat{J}$,
i.e. the eigenvalues of $\hat{J}^{\dagger}\hat{J}$. The latter are
 described by the so-called Laguerre ensemble
which independently emerged in
the description of quantum transport in mesoscopic
systems \cite{SN}, see e.g.\cite{RMT2} and references therein.

 All these facts make the
theory of random Hermitian matrices of various kinds a powerful and
versatile
tool of research in different branches of modern theoretical physics
\cite{RMT1,RMT2}.

Very recently complex eigenvalues of non-Hermitian random matrices have
also attracted much research interest due to their relevance to
several branches of theoretical physics.
Most obvious motivation comes from the quantum
description of {\it open} (or scattering)
quantum chaotic systems \cite{VWZ,LW,FSR,EfHam1} whose fragments can
escape, at a given energy, to infinity or come from infinity.
For systems of this kind  the notion of
discrete energy levels loses its validity. Actually, chaotic scattering
manifests itself in terms of a high density of  poles of
the scattering matrix placed
irregularly in the complex energy plane. Each of these poles, or {\it
resonances}, $ {\cal E}_{k}=E_{k}-\frac{i}{2}\Gamma_{k}$,
is characterized not only by energy
$E_{k}$  but also by a finite width $\Gamma_{k}$
defined as (twice) the imaginary part of the corresponding complex energy and
reflecting the finite lifetime of the states in the open system.
Naturally, such a picture stimulated various groups to develop
a statistical description of resonance parameters
\cite{Sokr,Sok,FS,reso,numr}.

Recently, the progress in numerical techniques and
computational facilities made available high accuracy patterns of
resonance poles for realistic atomic and
molecular systems in the regime of quantum chaos,
see e.g. \cite{Main,Blumel,Man,Greb} as well as for such model
quantum chaotic systems as quantum graphs \cite{graphs} and
a quantum particle under time- and space-periodic
perturbation \cite{KolRev}.
Due to the presence of   resonances the elements of the scattering matrix
 show irregular fluctuations as functions of the energy of incoming
waves, see \cite{Smilansky} and references therein.
The main goal of the theory of quantum chaotic scattering is to
provide an adequate statistical description of such a behaviour.

The most natural framework
for incorporating a random matrix description of the chaotic scattering
(and hence for addressing statistics of resonances) is the so-called
"Heidelberg approach" suggested in the pioneering paper \cite{VWZ}
and described in much detail in \cite{FSR,EfHam1}.
The starting point of that approach is
a particular formulation of the scattering theory
based on the notion of an "effective Hamiltonian". Such formulation
goes back to the ideas introduced in the context of Nuclear Physics
in classical works by Kapur and Peierls, Wigner, Bloch, Feshbach
and others. It was further developed, refined and clearly presented
 in the book by
Mahaux and Weidenm\"{u}ller \cite{Mah}, see also recent discussions
in \cite{EfHam1,EfHam2}. Following that approach
one introduces
a division of the Hilbert space of the
scattering system into two parts: the
"interaction region" and the "channel region". The channel region
is supposed to describe a situation of two fragments being apart
far enough to neglect any interaction between them. Under these
conditions their motion along the collision coordinate is
described by a superposition of incoming and outgoing plane waves
with wavevectors depending on the internal quantum states of the
fragments. We assume that at given energy $E$ exactly $M$ different
quantum states of the fragments are allowed, defining $M$
"scattering channels" numbered by the index $a$. At the same time, the second part of the Hilbert space
is to describe the situation when fragments are close to one
another and interact strongly. Correspondingly, any wavefunction
of the system can be represented as a two-component
vector, with first (second) component describing the
wave function inside the interaction (channel) region.

Using the standard
 methods of scattering theory exposed in detail in \cite{FSR,EfHam1,EfHam2}
one can relate
the two parts of the wave function to one another
and finally arrive at
the following representation of the
energy-dependent scattering matrix $\hat{S}$ in terms
of an effective non-Hermitian Hamiltonian
${\cal H}_{eff}=\hat{H}-i\hat{\Gamma}$:
\begin{equation}\label{def}\begin{array}{c}
\displaystyle{S_{ab}(E)=\delta_{ab}-2i\pi\sum_{ij}
W^*_{ai}[E-{\cal H}_{eff}]_{ij}^{-1}W_{jb}}
\end{array}\end{equation}
with the Hermitian Hamiltonian  $\hat{H}$ describing the {\it closed}
counterpart of the open system
(i.e. interaction region decoupled from the channel one)
and the anti-Hermitian part $i\hat{\Gamma}$
arising due to a coupling to open scattering channels. In this
expression
 the Hamiltonian $\hat{H}$ is written in some arbitrary basis
of states $|i\rangle$, such that
$H_{ij}=\langle i|\hat{H}|j\rangle$.
 The amplitudes $W_{ai},\quad a=1,2,...,M$ are
matrix elements coupling the internal motion in an "internal"
state $|i\rangle$ to one out of M open channels $a$. One also has to
choose the anti-Hermitian part to be
$ \Gamma_{ij}=\pi\sum_{a}W_{ia}W_{ja}^{*}$ in order to ensure the
unitarity
of the $M\times M$ scattering matrix $\hat{S}(E)$ \cite{Mah}.
The eigenvalues ${\cal E}_k$ of the operator ${\cal H}_{eff}=\hat{H}-i\hat{\Gamma}$
 are poles of the scattering matrix and
have a physical interpretation as
{\it resonances}: long-lived intermediate states to which discrete
energy levels of the closed system are transformed due
to coupling to continua. In particular, one can show that\cite{FSR}
\begin{equation}\label{dets1}
\det{S(E)}=\frac{\det{\left(E-{\cal H}_{eff}^{\dagger}\right)}}
{\det{\left(E-{\cal H}_{eff}\right)}}=\prod_{k=1}^N
\frac{\left(E-{\cal E}_k^*\right)}{\left(E-{\cal E}_k\right)}
\end{equation}
expressing the total scattering phaseshift $\phi(E)=\log{\det{S(E)}}$
in terms of the egenvalues ${\cal E}_k$.

Applying these constructions to the quantum chaotic scattering
it is natural to expect that universal properties of
scattering systems are inherited from the corresponding
universality of the energy levels of their closed counterparts.
Thus, we expect the scattering characteristics
(in particular, the statistics
of resonances and phaseshifts) could be adequately described by
 a random matrix approximation on a scale
comparable with the typical level spacing $\Delta$.
In contrast, a smooth energy
dependence of $S$-matrix elements on a
much larger energy scale must be system-specific.

According to the general idea one incorporates the random matrix
description of  quantum chaotic systems by replacing the Hamiltonian
$\hat{H}$ by a random matrix of appropriate symmetry. Namely,
chaotic systems with preserved time-reversal invariance (TRI) should be
described by matrices $H_{ij}$ which are real symmetric. On the other hand
systems with broken TRI are modelled by complex Hermitian matrices from
the Gaussian Unitary Ensemble (GUE) \cite{RMT1}.
The last step of the Heidelberg approach
is performing the ensemble averaging non-perturbatively.
Usually it is done in the framework of the so-called
supersymmetry technique introduced initially by
Efetov \cite{Efbook} and adopted to the description of quantum chaotic scattering by Verbaarschot, Weidenm\"{u}ller and Zirnbauer \cite{VWZ}.

The Heidelberg approach turns out to be a very powerful tool for
extracting different universal properties of open chaotic systems, and many important quantities characterizing the chaotic scattering
were successfully investigated along these lines \cite{LW,FSR,EfHam1,Gor}.
In particular, analytical predictions of the distribution of resonance
widths \cite{FSR,FS} were found to be in good agreement with available
data obtained numerically for realistic models of chaotic
quantum scattering \cite{graphs,KolRev,Seba}.

In fact, the outlined description of the wave scattering
can be looked at as an integral part
of the general theory of linear dynamic open systems in terms of the
input-output approach. These ideas and relations
 were developed in system theory and
engineering mathematics many years ago, see e.g. \cite{engmat}
 going back to the pioneering works by M. Liv\v{s}ic \cite{Liv}.
 Liv\v{s}ic himself stressed
equivalence of his mathematical constructions to those
used by nuclear physicists, but that
 theory and subsequent developments went practically
 unnoticed by the quantum-chaos community. A brief description of the
main constructions and interpretations of 
the linear open systems approach (in particular, a short derivation of the
equation Eq.(\ref{def})) can be found in \cite{FSJETP} and is not repeated here.
Let us only mention that the unitary scattering matrix $\hat{S}(E)$ 
is known in the mathematical
literature as the characteristic matrix-function of the non-Hermitian
fundamental operator ${\cal H}_{eff}$. 

In the scattering theory leading to Eq.(\ref{def})
 the time was considered to be a continuous parameter.
On the other hand, a very useful instrument in the analysis of classical
Hamiltonian systems with chaotic dynamics are the so-called area-preserving
 chaotic maps \cite{Sm2,Haakebook}. They appear naturally either as a mapping
of the Poincar\'{e} section onto itself, or as result of a stroboscopic
description of Hamiltonians which are periodic functions of time.
Their quantum mechanical analogues are unitary operators which act on
Hilbert spaces of finite large dimension $N$, and are often referred
to as evolution, scattering or Floquet operators, depending on
the given physical context. Their eigenvalues consist of
$N$ points on the unit circle (eigenphases). Numerical studies of
various classically chaotic systems suggest that the eigenphases
conform statistically quite accurately the results obtained for
unitary random matrices of a particular symmetry (Dyson circular ensembles).

Let us now imagine that a system represented by
 a chaotic map ("inner world") is embedded in a
larger physical system ("outer world") in such a way
that it describes particles which can
come inside the region of chaotic motion and leave
it after some time via $M$ open channels. Models of such type
appeared, for example, in \cite{BGS1}.

The general linear system theory \cite{engmat} considers
dynamical systems with discrete time as frequently as
those with continuous time. The corresponding construction is
discussed in \cite{FSJETP} and we mention here only its gross features.
 For a closed linear system characterized by a wavefunction $\Psi$
 the "stroboscopic" dynamics amounts to a 
linear unitary map $\Psi(n)\to\Psi(n+1)$, such that 
$\Psi(n+1)=\hat{u}\Psi(n)$. The unitary evolution 
operator $\hat{u}$ describes
the "closed" inner state domain decoupled both
from input and output spaces. Then a coupling that makes the system
open must convert the evolution operator $u$ to a contractive 
operator $\hat{A}$ such that ${\bf 1}-\hat{A}^{\dagger}\hat{A}\ge 0$.
It is easy to understand that one can always choose $\hat{A}=\hat{u}
\sqrt{{\bf 1}-\hat{\tau}\hat{\tau}^{\dagger}}$ 
where the matrix $\hat{\tau}$ is a rectangular 
$N\times M$ diagonal with the entries
$\tau_{ij}=\delta_{ij}\tau_j\, , \, 1\le i\le N\, ,\,  
{1\le j\le M} \ ({\rm for } M \le N ),\quad 0\le
\ \tau_j\le 1$. Such a form  automatically
ensures ${\bf 1}-\hat{A}^{\dagger}\hat{A}=\tau\tau^{\dagger}\ge 0$
and suggests a clear interpretation of the constituents of the model.
Namely, the equation $\Psi(n+1)=\hat{A}\Psi(n)$ describes an irreversible
decay of any initial state $\Psi(0)\ne 0$ when an input signal is absent.
On the other hand, assuming a nonzero input and zero
initial state $\Psi(0)=0$ one can relate 
the (discrete) Fourier-transforms of the input and output signals
at a frequency $\omega$ to each other by a $ M\times M$ {\it unitary}
 matrix $\hat{S}(\omega)$ given by
\begin{equation} \label{8}
\hat{S}(\omega)=\sqrt{{\bf 1}-\hat{\tau}^{\dagger}\hat{\tau}}
-\hat{\tau}^{\dagger}
\frac{1}{e^{-i\omega}{\bf 1}-\hat{A}}\hat{u}\hat{\tau}\ .
\end{equation}
This equation is a complete discrete-time analogue of Eq.(\ref{def}).
In particular, using the idenity $\det{\left(1-\hat{A}_1\hat{A}_2\right)}=
\det{\left(1-\hat{A}_2\hat{A}_1\right)}$ valid for two arbitrary matrices
$\hat{A}_{1,2}$ as well as the relation 
$\frac{1}{\sqrt{{\bf 1}-\hat{\tau}^{\dagger}\hat{\tau}}}\tau^{\dagger}
=\tau^{\dagger}\frac{1}{\sqrt{{\bf 1}-\hat{\tau}\hat{\tau}^{\dagger}}}$
it is straightforward to verify that
\begin{equation}\label{dets2}
\det{\hat{S}(\omega)}=e^{-i\omega N}
\frac{\det{\left(\hat{A}^{\dagger}-e^{i\omega}\right)}}
{\det{\left(e^{-i\omega}-\hat{A}\right)}}
= e^{-i\omega N}\prod_{k=1}^N
\frac{\left(z_k^{*}-e^{i\omega}\right)}{\left(e^{-i\omega}-z_k\right)},
\end{equation}
where $z_k$ stand for the complex eigenvalues of the matrix $\hat{A}$
which all belong to the interior of the unit circle: $|z|<1$. 
This relation is an obvious analogue of Eq.(\ref{dets1}) and
gives another indication that $z_n$ indeed play 
the role of resonances for the discrete time systems.

Now assume that the motion outside the inner region is regular.
According to the general strategy one 
should be able to describe generic features
of quantized maps representing chaotic inner dynamics by
 choosing the matrix $\hat{u}$ to be a member
of the Dyson circular ensemble of random matrices.
Then one finds: $\hat{\tau}^{\dagger}\hat{\tau}= {\bf 1}- |\overline{\hat{S}
(\omega)} |^2  $, with the bar standing for the averaging of
$\hat{S}(\omega)$ in Eq.(\ref{8}) over $\hat{u}$.
 Comparing this result with \cite{LW,FS} we see that
the $M$ eigenvalues $0\le T_a\le 1$ of the $M\times M$ matrix
$\hat{\tau}^{\dagger}\hat{\tau}$ play the role of the so-called transmission coefficients and describe a
particular way the chaotic region is coupled to the outer world.

This line of reasoning was motivated by recent papers reviewed in
\cite{KolRev}. The authors considered the Floquet
description of a Bloch particle in a constant force and periodic
driving. After some approximations the evolution of the system is
described by a mapping: ${\bf c}_{n+1}={\bf F}{\bf c}_{n}$, where
the unitary Floquet operator ${\bf F}=\hat{S}\hat{U}$ is the product of a
unitary "M-shift"
$\hat{S}: S_{kl}=\delta_{l,k-M}, \, l,k=-\infty,...,\infty$
and a unitary matrix $\hat{U}$. The latter is
effectively of the form
$\hat{U}=\mbox{diag}(\hat{d_1},\hat{u},d_2)$, where $\hat{d}_{1,2}$
are (semi)infinite diagonal matrices
 and $\hat{u}$ can be
taken from the ensemble of random $N\times N$ unitary matrices.
One can check that such a dynamics can easily be brought to the standard
form discussed above, with all
$M$ diagonal elements of the $N\times M$ matrix $\tau$ are equal to
unity\footnote{Actually, the original paper \cite{KolRev} employed a slightly
different but equivalent construction dealing with
an "enlarged" internal space of the dimension $N+M$. We prefer to
follow the general scheme because of its conceptual clarity.}.

A direct inspection immediately shows that the non-vanishing
 eigenvalues of the fundamental operator $\hat{A}$ for the 
latter example  coincide
with those of a $(N-M)\times (N-M)$ subblock of the  random
unitary matrix $\hat u$. Complex eigenvalues of such "truncations" of random
unitary matrices were studied analytically by
{\. Z}yczkowski and Sommers in \cite{KS}, and we reproduce their
study in Sec.{\it 3.1} .

In fact, truncated unitary
matrices represent only a particular case of random
contractions $\hat{A}$. The ensemble of general $N\times N$ random contractions
$\hat{A}=\hat{u}
\sqrt{{\bf 1}-\hat{\tau}\hat{\tau}^{\dagger}}$ describing a chaotic map with broken
time-reversal symmetry
can be shown to have the following probability measure in the matrix space:
\begin{equation}\label{0}
{\cal P}(\hat{A})
d\hat{A}\propto \delta(\hat{A}^{\dagger}\hat{A}-\hat{G}) d\hat{A}\quad,\quad
\hat{G}\equiv {\bf 1} -\hat{\tau}\hat{\tau}^{\dagger}
\end{equation}
where $d\hat{A}=\prod_{ij} d{\re A}_{ij}d{\im A}_{ij}$ and the normalization constant is independent of $\hat G$. It follows by averaging the matrix $\delta$-function $\delta( \hat{A}-\hat{u} \sqrt{{\bf 1}-\hat{\tau}\hat{\tau}^{\dagger}})$  over Dyson's circular unitary ensemble.
The $N\times N$ matrix $\hat{\tau}\hat{\tau}^{\dagger}={\bf 1}-
\hat{G}\ge 0$  is natural to call the deviation matrix
and we denote it $\hat{T}$.
It has $M$ nonzero eigenvalues coinciding
 with the transmission coefficients $T_a$ introduced above.
The particular choice $T_{i(\le M)}=1,\,\, T_{i(> M)}=0$
corresponds to the case considered in \cite{KS}.
Eq. (\ref{0}) describes an ensemble of subunitary matrices ${\hat A}$ with given eigenvalues of ${\hat A}^{\dagger}{\hat A}$, i.e. with specified singular values of the matrix ${\hat A}$. The question of characterizing the locus of complex eigenvalues for a matrix with prescribed singular values was considered in classical
papers by Horn and Weyl \cite{sing}. In essence, we provide a statistical answer to this question. Indeed, rotating the matrices ${\hat A}$ by a general unitary transformation and averaging over the unitary group yields a natural measure on matrices ${\hat A}$ with given singular values. In section {\bf 3} we will derive the corresponding joint density of eigenvalues and their correlation functions.\\

Although the initial interest in non-Hermitian and non-unitary
deformations of random matrices
had its origin in the theory of quantum chaotic scattering,
soon after it got a much of stimulus from the QCD applications of
chiral random matrices. The reason is that in order to
 describe Dirac eigenvalues in the
presence of non-zero chemical potential $\mu$ one has to replace the
(anti-)Hermitian Euclidean Dirac operator, see Eq.(\ref{dirac}),
with an operator of the following structure:
\begin{equation}\label{dirmu}
\hat{D}_{\mu}=\left(\begin{array}{cc}{\bf 0}&i\hat{J}+\mu{{\bf 1}}\\
i\hat{J}^{\dagger}+\mu{{\bf 1}}&
{\bf 0} \end{array}\right)
\end{equation}
Such matrices are no longer (anti-)Hermitian and their eigenvalues
go away from the (imaginary) axis to the complex plane.
This phenomenon clearly bears some similarity to the effect of opening
a closed system by converting its eigenvalues to resonances.
As the result, the fermion determinant in the partition function
(\ref{partf}) has a complex phase. On the level of true lattice
simulations such a phase results in major computational difficulties.
All this makes random matrix models a very useful tool for getting
insights into gross universal features of the
analytical structure of the QCD partition function. In turn,
information
on such a structure is based on the knowledge of statistics
of complex eigenvalues of the Dirac operator with non-vanishing
chemical potential
 which stimulated a lot of research of such eigenvalues
 \cite{QCD1,QCD,QCD2,QCD3,QCD4,QCD5,QCD6}.

Finally, it deserves to be briefly mentioned
that objects related to weakly non-Hermitian
random matrices emerge also in the context of interesting problems of
 motion and localization of a quantum particle in
disordered media subject to an {\it imaginary} vector potential
 \cite{Efnonh} or  a random scalar potential \cite{IzSi}.
The first of these problems is related to
the motion of superconductor
flux lines in
a sample with columnar defects,
the second
to the motion of a self-repelling polymer chain and other
interesting problems (see the cited references).
 These developments clearly show that, apart from
being a rich and largely unexplored mathematical object,
weak non-Hermitian deformations of random matrices
enjoy direct physical applications
and deserve to be studied in more detail.

 The goal of the present paper is to give an overview of
results and methods for dealing with spectra and eigenvectors of 
various non-Hermitian and non-unitary deformations of 
classical ensembles of random
matrices. For the reasons which should be clear from our discussion above
we concentrate on the regime
of {\it weak non-Hermiticity} recognized in \cite{FKS1}. It occurs when
the imaginary parts of typical eigenvalues are comparable
with the mean {\it separation} between
neighbouring eigenvalues along the real axis. A simple analysis
shows that the last condition is satisfied when the
traces of the Hermitian matrix $H$ and the perturbing anti-Hermitian
matrix $i\Gamma$ are related as: $\lim_{N\to \infty}
\frac{ N \mbox{Tr}\Gamma^2}{\mbox{Tr} H^2}<\infty$. 
We do not discuss at all interesting questions related to strongly
non-Hermitian matrices, see e.g. papers \cite{strong} and references 
therein.

The first weakly non-Hermitian ensemble analysed in full generality
\cite{FKS3} was that
 of non-Hermitian {\it Gaussian} deformations,
  with the anti-Hermitian part taken independently from the GUE.
For this case
 one can develop
a rigorous mathematical theory based on the method of
orthogonal polynomials. Such a method allows one to study
correlation properties of complex spectra to the same degree as
is typical for earlier studied classes of random matrices \cite{Mehta}.
We briefly discuss the results obtained in this model in subsection {\it 2.1}
(the detailed exposition of the method and the derivation of the
results can be found in \cite{FKS3}) as well as recent progress
for other  Gaussian ensembles and closely related models.
 We also introduce an
alternative way of addressing the problem which places the
correlation function of spectral determinants as primary object.
 This second method turns out to be
indispensable when one deals with
 {\it non-Gaussian} deformations
as those emerging in the theory of chaotic scattering.
The latter problem was successfully studied in the work by Fyodorov
and Khoruzhenko \cite{FK}
and we discuss in subsection {\it 2.2} the main ingredients of their solution.
At the end of that subsection we indicate a close similarity between
 the results obtained for the Gaussian and for the finite-rank
deformations.

Unfortunately, it is not known at the moment how to extend
the exact methods to the case of
non-Hermitian deformations of real symmetric matrices.
The last question is of great practical
interest for the application of the obtained results
to chaotic scattering. Indeed, the majority of
quantum chaotic
systems possesses time-reversal invariance and hence should be
adequately
described by random matrices which are real symmetric \cite{RMT1}.
Restricting our attention to the simplest,
still highly non-trivial statistical measure -
the mean density of complex eigenvalues of the resulting complex symmetric effective Hamiltonian ${\cal H}_{eff}$ -the answer can be given
in a closed form \cite{SFT}. The object turns out to be
accessible via Efetov's supersymmetry approach which we
 shortly discuss in subsection {\it 2.3}.

Very recently the properties of bi-orthogonal eigenvectors
of non-Hermitian random matrices attracted considerable interest
of various researchers
\cite{CM,Schom,Janik,MS,FM}. In particular, an essential boost to these
studies was given by the demonstrated relevance of the non-Hermitian
effective Hamiltonian ${\cal H}_{eff}=\hat{H}-i\hat{\Gamma}$ to the problem of lasing in random media \cite{Schom,Haake,Hacken}. Indeed,
for systems of such a type the non-orthogonality of eigenmodes is known to play an important role \cite{Pet1}.
Similar effects were discussed in relation to the unimolecular decay in $NO_2$ molecules \cite{Greb}.
In section {\bf 4} we discuss this issue for the regime of weak non-Hermiticity.

Most of our own results discussed in this review were reported
earlier in the form of short communications \cite{KS,FK,SFT,FM,FTS,prel,FSJETP}
and in an unpublished paper by one of the authors \cite{unpub}.
Some (but by far not all) details of the
underlying derivations are elucidated in the present text, as far as space restrictions allow. In particular,
  appendix A contains the derivation of the formulas
related to various aspects of the Itzykson-Zuber-Harish-Chandra integration over the unitary group extensively used in both \cite{FK} and \cite{FSJETP}.

\section{Non-Hermitian Deformations of Random Matrices}
\subsection{Gaussian deformations: from Wigner-Dyson
to Ginibre eigenvalue statistics.}

In the present section we concentrate on a particular
case of almost-Hermitian random matrices with i.i.d. entries
$\hat{J}=\hat{H}+iv\hat{A}$
 where $\hat H $ and
$\hat A$ are taken independently from the
Gaussian Unitary Ensemble (GUE) of {\it Hermitian } matrices $\hat X = \hat X^\dagger$
with   probability density
${\cal P}(\hat{X})=Q_N^{-1}
\exp{\Big(-N/2J_0^{2}\ \mbox{Tr}\ \hat{X}^2 \Big)}$
 .

Let us now introduce a new parameter $\tau=(1-v^2)/(1+v^2)$
and choose the scale constant $J_0^2$ to be equal to
$(1+\tau)/2$, for the sake of convenience.
The parameter $\tau$ controls the magnitude of
correlation between
$J_{jk}$ and $J_{kj}$: $\langle J_{jk}J_{kj} \rangle = \tau /N $,
hence
the degree of non-Hermiticity.
This is easily seen from the probability density function
for our ensemble of the random matrices $\hat{J}$:
\begin{eqnarray}
{\cal P}(\hat{J},\hat{J}^{\dagger})
=C_N^{-1} \exp \left (-\frac{N}{(1-\tau^2)} \ \mbox{Tr}
(\hat{J}\hat{J}^\dagger -\tau \ \re\  \hat{J}^2) \right ),
\label{P(J)}
\end{eqnarray}
where $C_N=[\pi^2(1-\tau^2)/N^2]^{N^2/2}$.
All the $J_{jk}$  have zero mean and variance
$\langle |J_{jk}|^2 \rangle =1/N$ and only
$J_{jk}$ and $J_{kj}$ are pairwise correlated.
If $\tau =0$ all the $J_{jk}$
are mutually independent and
we have maximum non-Hermiticity.
When $\tau $ approaches unity, $J_{jk}$ and
$J_{kj}^*$ are related via $J_{jk}=J_{kj}^*$
and we are back to an ensemble of Hermitian matrices.

Our first goal is to determine the $n$-eigenvalue correlation functions
in the ensemble of random matrices specified by Eq.\ (\ref{P(J)}).
The density of the joint distribution of
eigenvalues in this ensemble is given by
\begin{eqnarray}\label{P(Z)}
\fl {\cal P}_N(\{Z\})=c_N(\tau)|\Delta(\{Z\})|^2
\exp \left (\frac{-N}{1-\tau^2} \sum_{j=1}^N
\Big[|Z_j|^2 - \frac{\tau}{2}(Z_j^2+{Z_j^*}^2) \Big]  \right )\
\end{eqnarray}
where $c_N(\tau)= \frac{N^{N(N+1)/2}}{\pi^N 1!
\cdots N! (1-\tau^2)^{N/2}}$, $\{Z\}=(Z_1, \ldots, Z_N)$
is the vector of eigenvalues
and $\Delta(\{Z\})=\prod_{j<k}(Z_j-Z_k)$
is the corresponding Vandermonde determinant.

To derive Eq.~(\ref{P(Z)}) we integrate
${\cal P}(\hat{J},\hat{J}^{\dagger})$ from Eq.~(\ref{P(J)})
over the surface of all complex matrices
whose eigenvalues are $Z_1, \ldots Z_N$.
Following Dyson (\cite{Mehta},
p.501, see also \cite{Edel}) we decompose
every complex matrix with distinct eigenvalues as
$ \hat{J}=\hat{U}(\hat{Z}+\hat{R})\hat{U}^{\dagger}$,
where $\hat{Z}=\mbox{diag}\{ Z_1, \ldots Z_N \}$,
$\hat {U}$ is  a unitary matrix, and
$\hat{R}$ is a strictly upper-triangular one.
If we label the eigenvalues and
require the first
non-zero element in each column of $\hat{U}$ to be positive, then
the decomposition is unique (it is frequently referred to as "Schur decomposition").
The Jacobian of the transformation $\hat{J}\to \{\hat{Z},\hat{R}
,\hat{U}\}$
depends only on $\hat{Z}$ and is given by
the squared modulus of the Vandermonde determinant. So, integrating out
$\hat R$ and $\hat U$ is straightforward and the resulting expression
is Eq.~(\ref{P(Z)}).

The correlations between eigenvalues are conventionally characterized
by the $n$-eigenvalue correlation functions. To define them
we subdivide the vector of eigenvalues $\{ Z \}=(Z_1,...,Z_N)$ into
two
parts $\{ z\}=(Z_1,...,Z_n)$ and $\{ \xi\}=(\xi_1,...,
\xi_{k},...,\xi_{N-n})$ identifying $\xi_{k}=Z_{n+k}$.
 Then the correlation functions
are given by:
\begin{eqnarray}\label{corfunR}
R_n(\{z\})=\frac{N!}{(N-n)!}\int d^2\{\xi\}\
{\cal P}_N\{\{z\}, \{\xi\}\}
\label{R_n}
\end{eqnarray}
where ${
d^2\{\xi\}}\!=\!\prod_{l=1}^{N-n} d\re \xi_l\ d\im \xi_l$.

The form of the distribution
Eq.~(\ref{P(Z)}) allows one to employ
the powerful method of orthogonal polynomials \cite{Mehta}.
Let $H_n(z)$ denote the
$n$-th Hermite polynomial,
\begin{eqnarray} \label{H}
H_n(z)=\frac{(\pm i)^n}
{\sqrt{2\pi}}\! \exp{\left(\! \frac{z^2}{2}\! \right)}
\int_{-\infty}^{\infty}\!
dt\  t^n \exp{\left(-\frac{t^2}{2}\mp izt\right)}.
\end{eqnarray}
The
crucial observation borrowed from the paper \cite{orth}
(see also the related paper \cite{FJ})
 is that the polynomials
\begin{eqnarray}\label{p_n}
p_n(Z)=\frac{\tau^{n/2} \sqrt{N} }
{\sqrt{\pi }\sqrt{ n!}(1-\tau^2 )^{1/4}}
H_n\left( \sqrt{ \frac{N}{\tau}}Z\right),
\end{eqnarray}
$n=0,1,2, \ldots $,
are orthogonal in the {\it complex plane}
$Z=X+iY$
with the weight function
\[
w^2(Z)=\exp{\left (-\frac{ N}{(1-\tau^2)}\left[|Z|^2 -
\frac{\tau}{2}(Z^2+{Z^*}^2) \right]\right)},
\]
i.~e. $
\int d^2Z p_n(Z)p_m(Z^*)w^2(Z)  = \delta_{nm}$, where
$d^2 Z = dX dY$.

The standard machinery of the method of orthogonal
polynomials \cite{Mehta}
yields the functions $R_n(\{z\})$
in the form
\begin{eqnarray}\label{rn1}
R_n(\{z\})&=&\det \left[ K_N(Z_j,Z_k^*)\right]|_{j,k=1,\dots ,n}\ ,
\end{eqnarray}
where the kernel $K_N(Z_1,Z_2^*)$ is given by
\begin{eqnarray}
K_N(Z_1,Z_2^*)&=w(Z_1)w(Z_2^*)&\sum_{n=0}^{N-1}p_n(Z_1)p_n(Z_2^*).
\label{K}
\end{eqnarray}

With Eqs.\ (\ref{p_n})--(\ref{K}) at hand, let us first examine
the regime of strong non-Hermiticity, i.e. the case when
$\lim_{N\to \infty} (1-\tau ) > 0$, i.e. $\tau < 1$ independent of $N$. In this regime
the averaged density of eigenvalues $N^{-1}R_1(Z)$ is asymptotically
zero outside the ellipse $[\mbox{Re} Z/(1+\tau )]^2
+[\mbox{Im} Z/(1-\tau )]^2 \le 1$. Inside the ellipse
$\lim_{N\to \infty} N^{-1}R_1(Z) = [\pi (1-\tau^2)]^{-1}$.
This sets a microscopic scale on which the averaged number of
eigenvalues
in any domain of unit area remains finite
when $N \to \infty $. Remarkably, the $\tau$-dependence is
essentially trivial on this scale: the statistical properties
of eigenvalues are described by $\tilde R_n (z_1, \ldots , z_n)
\equiv N^{-n} R_n (\frac{z_1}{\sqrt{N}}, \ldots ,
\frac{z_n}{\sqrt{N}} )$ and
\begin{eqnarray}\label{Gin}
 \fl \lim_{N\to \infty} \tilde R_n (z_1, \ldots , z_n)=
 \left[
\frac{1}{ \pi  (1-\tau^2)   }
\right]^n
e^{-
    \frac{1}{1-\tau^2}\sum_{j=1}^n |z_j|^2
   }
     \det
       \left[
       e^{
                   \frac{1}{1-\tau^2}z_jz_k^*
         }
     \right]|_{j,k=1,\dots ,n}\ .
\end{eqnarray}
This limiting relation can be inferred \cite{FKS3} from Mehler's
formula
for the Hermite polynomials \cite{S}.
After the trivial additional rescaling  $z \to z\sqrt{1-\tau^2} $
the expression on the right-hand side in  Eq.\ (\ref{Gin})
becomes identical to that found by Ginibre \cite{Gin}.

Now we move on to the regime of weak non-Hermiticity.
We know that in this regime
new non-trivial correlations occur
on the scale:
$\mbox{Im} Z_{1,2}=O(1/N)$, $\mbox{Re} Z_1-\mbox{Re}
 Z_2=O(1/N)$.
Correspondingly, we
introduce new variables $x,y_1,y_2,\omega$ in such a way that:
$x=\mbox{Re}\left(Z_1+Z_2)\right/\! 2$,
$y_{1,2}=N\mbox{Im}\left(Z_{1,2} \right)$,
$\omega=N\mbox{Re}\left(Z_1-Z_2\right)$,
 and consider them finite when performing the limit $N\to
\infty$.

Substituting Eq.(\ref{H}) into Eq.(\ref{K}) and
using the above definitions
we can explicitly perform the limit $N\to \infty$,
taking into account that $\lim_{N\to \infty} N(1-\tau)= \alpha^2/2$.
The details of the procedure are given elsewhere \cite{FKS3}.
In this regime
\begin{eqnarray}\label{kern}
\fl
\lim_{N\to \infty}\frac{1}{N^2}
K_N\left(x+\frac{\omega/2+iy_1}{N}, x-\frac{\omega/2-iy_2}{N}\right)=
\exp{\left\{- \frac{y_1^2+y_2^2}{\alpha^2}+\frac{ix (y_1-y_2)}{2}
\right\}}
\\
\lo \nonumber \times\frac{1}{\pi\alpha}
  \int_{-\pi\nu_{sc}(x)}^{\pi\nu_{sc}(x)}\frac{du}{\sqrt{2\pi}}
\exp{\left[-\frac{\alpha^2u^2}{2}-u(y_1+y_2)+i\omega u\right]},
\end{eqnarray}
with
$\nu_{sc}(X)=\frac{1}{2\pi}\sqrt{4-X^2}$ standing for the Wigner
semicircular density of
real eigenvalues of the Hermitian part $\hat{H}$ of the matrices
$\hat{J}$.

The kernel $K_N$ given by Eq.\ (\ref{kern})
determines all the properties of complex eigenvalues in the regime of
weak non-Hermiticity. For instance,
the mean value of the density
$\rho(Z)= \sum_{i=1}^N\delta^{(2)}(Z-Z_i)$
of complex eigenvalues $Z=X+iY$ is  given by $
\langle\rho(Z)\rangle= K_N(Z,Z^*)$, i.e. by
putting $y_1=y_2$ and $\omega=0$ in Eq.(\ref{kern})
 (cf. \cite{FKS1} found by the supersymmetry approach described in subsection {\it 2.3}).

One of the most informative statistical measures of the spectral
correlations
is the `connected' part of the two-point correlation function of
eigenvalue
densities:
\begin{eqnarray}\label{rr}
\left\langle \rho(Z_1)\rho(Z_2)\right\rangle_c
=\left\langle
\rho(Z_1)\right\rangle \delta^{(2)}(Z_1-Z_2)-{\cal Y}_2(Z_1,Z_2),
\end{eqnarray}
In particular, it determines the variance $\Sigma^2(D)=
\langle n(D)^2\rangle-\langle n(D)\rangle^2$ of the number
$n=\int_D d^2Z\rho(Z)$
of complex eigenvalues in any domain $D$ in the complex plane as:
\begin{eqnarray}
\Sigma_2(D)= \int_Dd^2Z_1\int_Dd^2Z_2\ [\langle\rho(Z_1)\rho(Z_2)\rangle-
\langle\rho(Z_1)\rangle\langle\rho(Z_2)\rangle]\\ \nonumber
=\int_Dd^2Z\ \langle\rho(Z)\rangle-\int_Dd^2Z_1\int_Dd^2Z_2\ 
{\cal Y}_2(Z_1,Z_2)
\end{eqnarray}

Comparing Eq.(\ref{rr}) with the definitions Eqs.\
(\ref{p_n}-\ref{K})
we see that
the {\it cluster function} ${\cal Y}_2(Z_1,Z_2)$
is expressed in terms of the kernel $K_N$
as ${\cal Y}_2(Z_1,Z_2)=\left|K_N(Z_1,Z^*_2)\right|^2$.

It is evident that in the limit of weak non-Hermiticity
the kernel $K_N$
depends on $X$ only via the semicircular density $\nu_{sc}(X)$.
Thus, it does not change with $X$ on the local scale comparable with
the mean spacing along the real axis $\Delta\sim 1/N$.

The cluster function is given by the following explicit expression:
\begin{eqnarray}\label{clexp}
\fl {\cal Y}(\omega,y_1,y_2)= \frac{N^4}{\pi^2 \alpha^2}
e^{-2\frac{y_1^2+y_2^2}{\alpha^2}}
\left|\int_{-\pi\nu(X)}^{\pi\nu(X)}\frac{du}{(2\pi)^{1/2}}
\exp{\left[-\frac{\alpha^2 u^2}{2}-u(y_1+y_2)+iu\omega\right]}\right|^2
\end{eqnarray}

The parameter
$a=\pi\nu(X)\alpha$ controls the deviation from Hermiticity
\footnote{ In our earlier Letter \cite{FKS3} we used the definition of
the
parameter $a$ different by a factor of 2 from the present one.}.
When $a\to 0$ the cluster function tends to GUE form
${\cal Y}_2(\omega,y_1,y_2)=\frac{N^4}{\pi^2}\delta(y_1)
\delta(y_2)\frac{\sin^2{\pi\nu(X)\omega}}{\omega^2}$.
In the opposite case
$a\gg 1$ we expect it to match with the regime of strong non-Hermiticity
emerging for $\alpha\sim N^{1/2}$.
To verify
this we notice that for large $\alpha^2$ the
integral in Eq.(\ref{clexp}) can be evaluated
by the saddle-point method. The saddle-point in that case
is $u_s=-(y_1+y_2-i\omega)/\alpha^2$ and to the leading order
the result is non-vanishing as long as
$|\mbox{Re} u_s|\le \pi\nu_{sc}(X)$ that is
$|y_1+y_2|\le \pi\nu_{sc}(X)\alpha^2$. Assuming that this is the case
one performs the Gaussian integration around $u_s$ and
finds (in the original variables $Z_1,Z_2$)
the expression equivalent (up to a trivial rescaling) to
that found by Ginibre \cite{Gin}:
${\cal Y}_2(Z_1,Z_2)=(N^2/\pi \alpha^2)^{2}
\exp\{-N^2|Z_1-Z_2|^2/{\alpha}^2\}$.
Remembering $\nu_{sc}(X)=\frac{1}{2\pi}\sqrt{4-X^2}$ and
$\alpha^2=Nv^2$
and performing the same procedure for the mean density
$\rho(Z)=K(Z,Z^*)$
one finds that
the condition for the saddle-point discussed above just ensures
that both $Z_1$ and $Z_2$ are inside an elliptic blob
$(\re Z)^2/4+(2\im Z)^2/v^2=1$
in the complex plane filled in uniformly with eigenvalues with the
density $\rho(Z)=N/(\pi v^2)$, in agreement with \cite{ellipse}. This is basically the
case of "strong" non-Hermiticity, see the discussion preceeding
Eq.(\ref{Gin}), and the "elliptic" law is believed to be the common feature for strongly non-Hermitian matrices with independent entries \cite{Edel,Bai}.

One can further calculate the Fourier transform
of the cluster function over its arguments $\omega,y_1,y_2$
and find an explicit expression for the spectral form-factor.
It allows one to determine the variance $\Sigma_2$ of a number of
eigenvalues in any domain $D$ of the complex plane. In this way
one traces, in particular, a tendency for gradual decorrelation
of real parts of the eigenvalues with growing non-Hermiticity.
Similarly, one can investigate various regimes of eigenvalue
repulsion in the complex plane by looking at small-distance behaviour
of the nearest neighbour distance distribution. All these
calculations are presented and thoroughly discussed in \cite{FKS3}.

Further calculations in the framework of the present model were done
 recently by Akemann \cite{QCD4}. Motivated by QCD applications
the author first suggested to
consider  the matrix integral
\begin{equation}\label{qcd3}
\fl {\cal Z}\left(\left\{m_f\right\}\right)=\int d\hat{J}d\hat{J}^{\dagger}\ 
{\cal P}({\hat J},{\hat J}^{\dagger})
\prod_{f=1}^{N_F}\left[\det{({\hat J}-im_f{\bf 1})}\det{({\hat J}^{\dagger}+im_f{\bf 1})}\right]
\end{equation}
with ${\cal P}({\hat J},{\hat J}^{\dagger})$ being the joint probability
density function of the random matrices $\hat{J}$, see Eq.(\ref{P(J)}).
This integral is a model for the Euclidean QCD partition function 
in 3 dimensions at nonzero chemical potential, 
when the chiral structure of the Dirac operator is irrelevant.
The latter fact gives some justification of the purely phenomenological model based on
Eq.(\ref{qcd3}).
On the other hand, the matrix integral is interesting by itself
and its calculation is very intimately related to calculating
the eigenvalue correlation function Eq.(\ref{R_n}).

Indeed, having in mind the decomposition
$(Z_1,...,Z_N)=(\{z\},\{\xi\})$ as in Eq.(\ref{R_n}) and
noticing that the Vandermonde determinant
$\Delta(\{Z\})$ can be rewritten as :
$$
\Delta(\{Z\})=\Delta(\{z\})\Delta(\{\xi\})\prod_{i=1}^n
\prod_{j=1}^{N-n}(Z_i-\xi_j)
$$
it is easy to see that the correlation function
$R_n(\{z\})$ as defined in Eq.({\ref{R_n}) can be put in the form:
\begin{equation}\label{example}
\fl R_n(\{z\})=\frac{c_n(\tau)}{c_{N-n}(\tau)}|\Delta(\{z\})|^2
\left[\prod_{i=1}^n w^2(Z_i)\right]\int d^2 \{\xi\}\
{\cal P}_{N-n}(\{\xi\})
\prod_{i=1}^n\left|\prod_{j=1}^{N-n}(Z_i-\xi_j)\right|^2
\end{equation}
where ${\cal P}_{N-n}(\{\xi\})$ is just the density of joint
probability of eigenvalues $\xi_1,....,\xi_{N-n}$ of a random Gaussian
non-Hermitian matrix $J_{N-n}=H_1+iv A$ of the size
$(N-n)\times(N-n)$. But the integral in the last equation is just
the expectation value of the   product of determinants:
$$
\left\langle\prod_{i=1}^n\left[\det\left(Z_i{\bf 1}-{\hat J}_{N-n}\right)
\det\left(Z_i^*{\bf
1}-{{\hat J} ^{\dagger}}_{N-n}\right)\right]\right\rangle_{J_{N-n}}
$$
Thus, we see that the problem of evaluating the multipoint
 correlation function $R_n(\{z\})$ of complex eigenvalues is, in
fact,
equivalent to calculating the RMT partition
function for QCD$_3$, Eq.(\ref{qcd3}).
Akemann\cite{QCD4} used this observation to express the partition
function ${\cal Z}$ in terms of the $R_n(\{z\})$, and
in this way exploited the orthogonal polynomial representation,
see Eqs.(\ref{rn1},\ref{K}).

Here we would like to note that by reversing this kind of
reasoning one could get access to the multipoint correlations
of complex eigenvalues if one were able to provide an independent
technique
of evaluating ${\cal Z}$. This was realized by Fyodorov and Khoruzhenko \cite{FK} who implemented such an idea
for a more general class of ensembles.
We will see below, that a reduction similar to that discussed above holds
 beyond the Gaussian case, provided one deals with the regime
of weak non-Hermiticity.
Moreover, it will be shown how to
perform the ensemble average  of the product of characteristic
polynomials for an {\it arbitrary fixed}
 non-Hermitian deformation  by mapping the analogue of
partition function Eq.(\ref{qcd3}) to a fermionic
version of a nonlinear $\sigma-$model.

Let us discuss briefly other types of (weakly) non-Hermitian
deformations based on the Gaussian case and studied by various
authors. Recently Akemann \cite{QCD5} considered a model with the
joint probability density of variables $Z_i$ defined by
\begin{eqnarray}\label{Pchir(Z)}
\fl{\cal P}^{(ch)}_N(\{Z\}) \propto |\Delta({\{Z\}}^2)|^2\,\,
|Z_1\ldots Z_N|^{2a+1}\exp \Big\{\frac{-N}{1-\tau^2} \sum_{j=1}^N
\Big[|Z_j|^2 - \frac{\tau}{2}(Z_j^2+{Z_j^*}^2) \Big]  \Big\}\ .
\end{eqnarray}
The new factors $|Z|^{2a+1}\,,\, a>-1$ as well as squared variables
inside the Vandermonde determinant were introduced in an ad hoc way
to make this ensemble a candidate for
chiral extension of the Gaussian case,
provided one interpretes the variables ${Z}$ as complex Dirac
eigenvalues. In such an interpretation  the parameter $a$ can be
related to the number of massless quark
flavours $N_F$, whereas the chemical potential is phenomenologically
identified with $\frac{1}{2}\sqrt{1-\tau^2}$.
The advantage of the model is that it still can be
solved by the orthogonal polynomial method, with (properly
normalized) Laguerre polynomials of complex argument
$L_k^{(a)}\left(\frac{Nz^2}{2\tau}\right)$ replacing the
Hermite polynomials. In particular, elegant explicit
expressions for the limiting form of the kernel
$K^{(a)}(Z_1,Z^*_2)$ were derived both in the strong non-Hermiticity
and
the weak non-Hermiticity regimes \cite{QCD5}.
On the other hand, the main drawback of the model
is its ad-hoc construction in terms of the variables $Z_i$.
In particular, it is completely
unclear how to construct the
corresponding random matrix model for the Dirac operator which could
have the starting formula
(\ref{Pchir(Z)}) as the joint probability density of
its eigenvalues. The hope is, nevertheless, that the universality
will again make the
eigenvalue correlations insensitive to the
particular choice of the underlying model \cite{QCD6}.
However, we believe that the full analysis of the random matrix
model Eq.(\ref{dirmu}) in the regime of weak non-Hermiticity
still remains an important unsolved problem.

In the spirit of the methods discussed in the present section
one can efficiently study
Gaussian ensembles corresponding to
the  weakly non-Hermitian {\it quaternion} random matrices.
Matrices with such (symplectic) symmetry are also important for QCD
applications \cite{QCD} and that fact inspired quite a few attempts
of understanding their properties
analytically \cite{EfKol,Hast,NK,Kanz}.
The most complete solution was presented by Kanzieper \cite{Kanz} who
used
the skew orthogonal polynomials.

Recently another interesting extension of the model Eq.(\ref{P(J)}) was
suggested
by Garcia-Garcia et al.\cite{GNV}. The authors introduced an ensemble
that interpolates between the GUE, the Ginibre ensemble of strongly
non-Hermitian matrices and the Hermitian Poisson ensemble with
uncorrelated
real eigenvalues. The physical motivation behind such a
generalization comes from the wish to understand better an interplay
between
the effects of non-Hermiticity and the effects of Anderson
localization which drive the statistics of real eigenvalues towards an
uncorrelated Poisson spectrum. It is claimed that the model can
adequately describe critical spectral statistics of open disordered
systems
close to the Anderson transition. The authors managed to
provide a closed
analytical expression for the eigenvalue correlations
in various regimes and concluded that critical statistics
is not modified by weak non-Hermiticity effects. They further put
forward an expectation that opening a critical system does not
affect the multifractal dimension of the wavefunction. Whether it is
indeed the case remains to be verified in numerical simulations.

\subsection{Finite-rank deformations of Hermitian matrices: the case
of chaotic scattering}
As it was discussed in the introduction, the  poles of the $S$-matrix
(resonances) are just the complex eigenvalues of an effective
random matrix Hamiltonian ${\cal H}_{eff}\!=\!\hat H\!-\!
i\hat \Gamma$. The requirement of the $S$-matrix unitarity and causality
restricts $\hat\Gamma$ to be of the rank $M$, characterized by $M$ positive eigenvalues $\gamma_c>0\,\, , c=1,\ldots,M$, the rest $N-M$ eigenvalues being identically zero.
To satisfy the
condition of weak non-Hermiticity
for the case of a few open channels: $N\gg M\sim 1$ means just to consider
these eigenvalues to be $N$-independent: $\gamma_c<\infty$.
The condition ensures the width $\Gamma_k$ of a
typical resonance to be comparable with the mean {\it separation} $\Delta$ between neighbouring resonances along the real axis.

Despite quite substantial efforts \cite{Sok,FS,FSR,SFT,FTS} non-perturbative results on resonance statistics for few-channel
scattering  were very restricted for a long time. The knowledge amounted
mainly to (i) the joint probability density of all resonances
for the system with a single open channel and Gaussian-distributed
transition amplitudes by Sokolov et al. $\hat W$ \cite{Sokr} and  (ii) the mean density
of $S$-matrix poles for arbitrary $M\!\ll\! N$ \cite{FSR,SFT}. A systematic analytical
approach to the statistical properties of resonances for systems with
broken time-reversal invariance was suggested
in \cite{FK}. In what follows we discuss that approach in more detail.
 
For the sake of notational convenience we have changed the sign in front of $ \Gamma$ \footnote{in the rest of this subsection we will denote matrices without "hat"}. This trivially amounts
to changing the signs of imaginary parts of all eigenvalues in the
resulting expressions. From now on we consider an ensemble of random $N\times N$ complex matrices ${J}={H}+i{\Gamma}$, where ${H} $ is $N\times N$ matrix taken from
a Gaussian Unitary Ensemble (GUE) of {\it Hermitian } matrices
with the probability density ${\cal P}({H})\propto
\exp{(-{N\over 2} \tr {H}^2)}$, ${H} = {H}^\dagger $. As for the
matrix ${\Gamma}=\Gamma^{\dagger}$, we consider it to be a {\it
fixed} nonnegative one: $\Gamma\ge 0$.

By analogy with the complex numbers,
\[ H=\frac{1}{2}(J+J^{\dagger})\equiv \mbox{Re} J;\;\;\; \Gamma
=\frac{1}{2i}(J-J^{\dagger})\equiv \mbox{Im} J \]
with the "only" difference that $\mbox{Re} J$ and $\mbox{Im} J$ do not commute.
Then the
probability density function in our ensemble of random matrices
${J}$ can be written in the form
\begin{equation}\label{P(J1)}
{\cal P}({J})\propto e^{-\frac{N}{2} \mbox{Tr}\, (\mbox{\small Re} J)^2}\delta
(\Gamma - \mbox{Im} J ).
\end{equation}
with a matrix $\delta$-function for Hermitian matrices.
We do not specify the multiplicative constant in Eq.(\ref{P(J1)}). It can be found from the normalization condition.
Similarly, and by the same reason, we will systematically
disregard multiplicative constants when dealing with probability
densities and correlation functions.

Eq.\ (\ref{P(J1)}) can be used to obtain the density of joint
distribution of eigenvalues by integrating ${\cal P}(J)$ over the
degrees of freedom that are complementary to the eigenvalues of
$J$. This again can be done following Dyson's method (see
preceding section).
To perform the integration over upper triangular matrices ${R}$ it is
technically convenient to use a Fourier-integral representation
for the $\delta-$function in Eq.\ (\ref{P(J1)}). This reduces the
corresponding integral to a Gaussian one and after
quite straightforward algebraic
manipulations the resulting expression is
\begin{equation}\label{P(Z1)}
{P}_N({Z})\propto e^{-\frac{N}{2}  \mbox{\small Re} \mbox{\small Tr} {Z}^2 -
\frac{N}{2}\mbox{\small Tr}
{\Gamma}^2 }|\Delta (\!{Z})|^2{Q}_M(\im {Z} ),
\end{equation}
where $ {Q}_M(\mbox{Im} {Z} ) $ is the remaining integral
\begin{equation}\label{y:1}
{Q}_M(\mbox{Im} {Z} )=\int\! {[dU]} \prod_{l=1}^N\delta
\left(\mbox{Im}{z}_l-({U}^{\dagger}{\Gamma}{U})_{ll}\right),
\end{equation}
over the unitary group $U(N)$, $[dU]$ being the Haar measure.
It is clear that we can consider $\Gamma$ to be diagonal, and
the integral Eq.(\ref{y:1}), has an obvious interpretation of the joint probability density of diagonal entries for a matrix with prescribed
set of eigenvalues ${\Gamma}$, randomly rotated by a unitary matrix ${U}$. Recently this interpretion was helpful in applications
to the problem of Wigner time-delay distribution \cite{SFS}.

To proceed further we need to integrate over $U$. Again it is
convenient to use the Fourier-integral representation for the
$\delta-$functions in Eq.\ (\ref{y:1}):
\begin{equation}\label{y:2}
{Q}_M(\mbox{Im} {Z} ) = \int\!\!\!\frac{dK}{(2\pi)^N}\
 e^{i\, \mbox{\small Im} \mbox{\small Tr}{K}{Z} }
\!\!\int\!{[dU]}\ e^{-i\, \mbox{\small Tr} {K} {U}^{\dagger}{\Gamma} {U} },
\end{equation}
where the first integration is over all real diagonal matrices
${K}$ of dimension $N$, $dK$ being $dk_1 \ldots dk_N$.

When the eigenvalues $\gamma_1,...,\gamma_N$
of ${\Gamma}$ are all \emph{distinct} the
integration over $U(N)$ can be performed using the famous
Itzykson-Zuber-Harish-Chandra (IZHC) formula \cite{IZHC}:
\begin{equation}\label{IZHC}
\int\!{[dU]}\ e^{-i\, \mbox{\small Tr} {K} {U}^{\dagger}{\Gamma} {U} } =
\!\!
\displaystyle{\frac{\det\left[e^{-ik_n\gamma_m}\right]_{n,m=1,...,N}}
{\Delta(-i\{K\})\Delta({\{\gamma\}})}}
\end{equation}
where $\{K\} =\diag (K_1,\ldots,K_N)$ and $\{\gamma\} =\diag (\gamma_1,\ldots,\gamma_N)$  and we have chosen the total volume of $U(N)$ equal to $1$.
 We, however, are mostly interested in the case when ${\Gamma}$ has a
small rank $M\ll N$, i.e.\  it has only $M$ nonzero eigenvalues
which we denote by $\gamma_1, \ldots , \gamma_M$, the rest $N-M$
being zero. This limit of
highly degenerate eigenvalues is difficult to perform in the
original IZHC formula.  A way to circumvent
this difficulty is discussed in the appendix A1 where it is shown that:
\begin{equation}\label{IZHCD}
\fl \int\!{[dU]}\ e^{-i\, \mbox{Tr} {K} {U}^{\dagger}{\Gamma} {U} }=i^{-N^2}
M!\frac{\det{\gamma}^{M-N}}{\Delta({\gamma})}
\int d\Lambda
\Delta(\Lambda)e^{-i\sum_{c=1}^M\gamma_c\lambda_c}
\prod_{j=1}^N\prod_{c=1}^M\frac{1}{\lambda_c-k_j}
\end{equation}
with $\Lambda =\diag (\lambda_1,\ldots,\lambda_M)$. Here it is meant that
the integration contour is chosen in such a way that $\mbox{Im}\,\lambda_c=0^{+}$.

To perform the integration over $K$ is a simple task with the use
of the identities:
$$
\prod_{c=1}^M\frac{1}{\lambda_c-k_j}=\sum_{c=1}^M\frac{1}{\lambda_c-k_j}
\prod_{s(\ne c)}\frac{1}{\lambda_c-\lambda_s}\ ;\
\int_{-\infty}^{\infty}{dk\over 2\pi} {e^{i\im z_j k}\over
 {\lambda_c-k}}=\left\{\begin{array}{cc}
-ie^{i\im z_j \lambda_c}&\im z_i>0\\0&\mbox{otherwise}\end{array}\right.
$$
Taking into account that due to
the specific structure of the matrices $J$ their eigenvalues lie
in the upper part of the complex plane we assume in all formulae below
 that $\im z_j > 0$ for all $j$.
This finally results in:
\begin{equation} \label{start}
\fl {Q}_M(\im Z)\propto M!\frac{\det {\gamma}^{M-N}}
{\Delta({\gamma})}\int_{R^M}\!\!d\Lambda\
e^{-i\sum_{c=1}^M\gamma_c\lambda_c}\Delta(\Lambda )\prod\limits_{j=1}^N
\sum_{q=1}^M\frac{e^{i\lambda_q \im z_j}} {\prod_{s(\ne
q)}(\lambda_q-\lambda_s)}\nonumber\ .
\end{equation}

The pair of equations (\ref{P(Z1)},\ref{start}) provides an
explicit representation for the joint probability density of $N$
resonances $z_j$ in the complex plane. For the particular case of
a non-Hermitian
rank-one perturbation of GUE matrices (single-channel case $M=1$) the density of joint probability of all complex eigenvalues was found
earlier by St\"{o}ckmann and Seba for a closely related model
of {\it random} couplings $\hat{W}$ \cite{Sokr}.
Their result follows from
our Eq.(\ref{P(Z1)},\ref{start}) by noticing (i) that for $M=1$
the $\lambda-$integration is trivial to perform, yielding
\begin{equation}\label{jpd}
{\cal P}_N(Z)|_{M=1}\propto e^{-\frac{N}{2} \mbox{Re} \mbox{Tr} Z^2}
|\Delta(Z)|^2\frac{e^{-N\gamma^2/2}}{\gamma^{N-1}}
\delta(\gamma-\sum_{j=1}^N \mbox{Im} z_j)
\end{equation}
and (ii) in their case the only nonzero eigenvalue $\gamma>0$ is a random variable with the probability density
${\cal P}(\gamma)\propto \gamma^{N-1}\exp{-\left[N\gamma/\gamma_0\right]}$.

Our main goal is to use equations (\ref{P(Z1)},\ref{start})
as a basis
for calculating the $n$-eigenvalue correlation functions
$R_n( \{z\})$ as defined in Eq.(\ref{R_n}).
In what follows we will calculate $R_n(\{z\})$ for arbitrary
fixed $n,M$ in the limit $N\to \infty$, but it may be instructive for the reader first to verify all the steps on the simplest case $M=1$,
starting from the Eq.(\ref{jpd}) and following \cite{FK}.
On the first stage we will
replace the integration over ${\xi}$ in (\ref{R_n}) by
averaging over the ensemble of
non-Hermitian random matrices $J_{N-n}=H_{N-n}+i\Gamma$,
 with $H_{N-n}$ being a GUE matrix of the
reduced size $(N-n)\times (N-n)$ (cf. Eq.(\ref{example})).
In this way one finds the following representation for the correlation functions:
\begin{equation}\label{interm}
\fl R_n(\{z\})\propto  \frac{{C}_{{\gamma}} (\{z\})}{\det
{\gamma}^n}\ |\Delta (\{z\})|^2\
e^{-\frac{N-n}{2}\sum_{j=1}^n \re\, z_j^2}
 \prod_{j=1}^n
 \sum_{k=1}^M\frac{e^{-2y_jg_k}}
{\prod_{s(\ne k)}(g_k-g_s)}
\end{equation}
where we introduced the notation $g_c=\frac{1}{2}(\gamma_c+\gamma_c^{-1})$
and denoted:
\begin{equation}\label{CC_}
{C}_{\tilde{{\gamma}}}(\{z\})=
 \left\langle\prod_{j=1}^n
\Big| \det
\left[z_j-{J}_{N-n}(\tilde{{\gamma}})\right]\Big|^2
\right\rangle_{GUE}
\end{equation}
The corresponding derivation is presented in appendix B and is valid
in the limit $n,M\ll N\to \infty$.

Thus, the problem amounts to evaluation of the correlation
function of the determinants in Eq.(\ref{CC_}). To proceed we first
write each of the determinants as a Gaussian integral over a set
of Grassmann variables $\chi,\chi^{\dagger}$:
\begin{equation}\label{Ber1}
\fl \prod_{p=1}^n\Big| \det
\left[z_p-{J}_{N-n}({\gamma})\right]\Big|^2=
\int d\chi\,d{\chi}^{\dagger}
e^{-\left({\chi}^{\dagger}\mbox{\bf Z}_{2n}\chi+{\chi}^{\dagger}
\mbox{\bf H}_{N-n}{\chi}+i{\chi}^{\dagger}\left[\Gamma\otimes\mbox{\bf  L}_{\,2n}\right]{\chi}\right)}\ .
\end{equation}

When this is done, the GUE average over $H_{N-n}$ becomes
trivial and yields the terms quartic with respect to the
Grassmannian vectors. These  terms can be further traded for an auxilliary
integration over a Hermitian matrix ${S}$ of the size $2n\times 2n$
(the so-called Hubbard-Stratonovich transformation).
Then the integration over the Grassmann fields becomes again Gaussian
and is trivially
performed yielding again a determinant.
Introducing diagonal matrices defined as:
$\mbox{\bf Z}_{2n} = \diag
(\{z\}, \{z\}^{\dagger})$ and $\mbox{\bf  L}_{\, 2n}= \diag( {\bf 1}_n,- {\bf 1}_n)$
and using the finite-rank
property of $\Gamma$ one easily brings the result
to the following form:
\begin{equation}\label{repr}
\fl {C}_{{\gamma}}(\{z\})=\int [d{S}]
 e^{-(N-n)\mbox{Tr}[\frac{1}{2}{S}^2-\ln(\mbox{\bf Z}_{2n}-i{S})]}
\prod_{c=1}^M \det\left[ \mbox{\bf 1}_{2n}+
i\gamma_c \mbox{\bf  L}_{\,2n} ( \mbox{\bf  Z}_{2n}- i{S})^{-1}\right]\ .
\end{equation}

 We are interested  in evaluating
the above integrals in the limit $N\to\infty$.
Let us now recall that nontrivial eigenvalue correlations are
expected to occur on such a scale when the eigenvalues are separated
by distances comparable with the mean eigenvalue separation for
GUE matrices $H$ (see preceding section), the latter distance
being of the order $(N-n)^{-1}$ for our
choice of ${\cal P}(H) $. Accordingly, it is convenient to separate
the "center of mass" coordinate $x=\frac{1}{n} \sum_{j=1}^n\re
z_j$ so that $z_j = x+\frac{\tilde {z}_j}{N-n}$,
where both the real and imaginary parts of $\tilde{z}_j$ are of the
order of unity in the limit when $N\to \infty $ and $M$ is
fixed. In this limit $R_n(\{z\})$ is effectively a function of
$ \{\tilde z\}$ ($x$ is fixed) which we are going to calculate.
To this end, we expand the logarithm in the exponent of
Eq.(\ref{repr}) as:
\begin{eqnarray}
\fl (N-n)\mbox{Tr}\ln(\mbox{\bf  Z}_{2n}-i{S})
\approx (N-n)\mbox{Tr}\ln(x  \mbox{\bf 1}_{2n}-i{S})+
\mbox{Tr}\left[\tilde{\mbox{\bf  Z}}_{2n}(x
\mbox{\bf 1}_{2n}-i{S})^{-1}\right]+...
\end{eqnarray}
keeping only terms up to the order $O(1)$ when $N\to\infty$.
With the same precision we set
$\tilde{\mbox{\bf Z}}_{2n}=x\mbox{\bf 1}_{2n}$
in the pre-exponential determinant
factor of the integrand, Eq.(\ref{repr}).

The next step towards evaluating
 the integral in (\ref{repr}) is to  diagonalize the Hermitian matrix
$S$ as: $S={U}_{2n}{\Sigma } U_{2n}^{-1}$, where ${U}_{2n}\in U(2n)$
and ${\Sigma }=\diag (\sigma_1,...,\sigma_{2n})$. Then, keeping
only the terms relevant in the limit $N\to \infty$, we obtain:
\begin{equation}\label{CC1}
{C}_{\gamma}(\{z\})\!=\!\int\!\! d\Sigma\,
\Delta^2(\Sigma) e^{-(N-n)\!\!\sum\limits_{k=1}^{2n}[\frac{
\sigma_k^2}{2}- \ln{(x-i\sigma_k)}]}\langle C({S})\rangle_{U(2n)}
\end{equation}
where
\begin{equation}
\fl \langle C({S})\rangle_{U(2n)}= \int [d U_{2n}]\,
e^{-\mbox{Tr}[\tilde{\mbox{\bf  Z}}_{2n}(x\mbox{\bf 1}_{\,2n}-i{S})^{-1}]}
\prod_{c=1}^M \det\big[\mbox{\bf 1}_{2n}+i\gamma_c\mbox{\bf  L}_{2n}(x\mbox{\bf 1}_{2n} - i{S})^{-1}\big]
\end{equation}

The form of the integrand in (\ref{CC1}) suggests exploiting the
saddle-point method in the integral over $\sigma_k$, $k=1,\ldots,
2n$.  Altogether there are $2^{2n}$ saddle-points:
$\sigma_k^{(s)}=-\frac{i}{2}(x+i\epsilon_k\sqrt{4-x^2})$ each
corresponding to a particular choice of $2n$ signs
$\epsilon_k=\pm 1$. It is easy to understand, however, that not all
these saddle points are equally important in the limit of large $N$.
The leading order contribution comes from
integration around  those saddle-points where exactly $n$
parameters $\epsilon_k$ equal 1 (the rest being equal -1). All
other choices can be neglected as they lead to lower order terms.
This is because of the presence of the Vandermonde determinant in
the integrand:
$$
\Delta^2(\Sigma)=\prod_{k_1<k_2}\left(\sigma_{k_1}-\sigma_{k_2}\right)^2
=\prod_{k_1<k_2}\left(\delta\sigma_{k_1}-\delta\sigma_{k_2}+
\frac{1}{2}(\epsilon_{k_1}-\epsilon_{k_2})\sqrt{4-x^2}\right)^2
$$
where we denoted by $\delta\sigma_{k}$ fluctuations around a given
saddle-point to be treated in the Gaussian approximation.
It is now evident, that we should require that the {\it minimum}
number of the differences
$\epsilon_{k_1}-\epsilon_{k_2},\quad k_1<k_2=1,2,...,n$
vanish. Otherwise, the Vandermonde determinant is proportional to
  higher powers of $\delta\sigma_{k}$, each power producing
a small factor $(N-n)^{-1}$ when integrating
around the saddle-point in the Gaussian approximation. This
requirement fixes the choice of the saddle-points described above.

 At the same time, all relevant saddle-points are {\it equivalent}
(because they can be transformed one to the other by an element of the
$U(2n)$ group) and
produce the same contribution. Therefore, we can take one of them,
 suppressing the multiplicative
constants as usual.
Our choice of $\Sigma^{(s)}$ is  $\Sigma^{(s)}=-\frac{i}{2}x
\mbox{\bf 1}_{2n}+\pi\nu(x)\mbox{\bf  L}_{2n}$,
where the symbol $\nu(x)$ stands for the semicircular density of real
eigenvalues of the matrices ${H}$,
$\nu(x)=\frac{1}{2\pi}\sqrt{4-x^2}$.
Substituting $\Sigma^{(s)}$ to the integrand it is easy to
verify that in the limit $N\to \infty$ we have:
\begin{equation}\label{CCC}
{C}_{{\gamma}}(\{z\})\propto
e^{\frac{N-n}{2}\sum_{j=1}^n\re z_j^2}
{C}^{s}_{{\gamma}}( {\{\tilde z\}})
\end{equation}
where $ \{\tilde z\}=(\tilde{z}_1, \ldots, \tilde{z}_n)$,
$\tilde{z}_j=(N-n)(z_j-x)$, $j=1,\ldots,n$,  and
\begin{equation}\label{c4}
\fl {C}^{s}_{{\gamma}}( {\{\tilde z\}})= \int [{\rm d}\mbox{\bf  Q}_{2n}]\  e^{-i\pi\nu(x)\mbox{Tr}\,
\tilde{\mbox{\bf  Z}}_{2n}\!\mbox{\bf  Q}_{2n} }
\prod_{c=1}^M\det \left[\mbox{\bf 1}_{2n}+
\frac{i\gamma_c x}{2}\mbox{\bf  L}_{2n}+ \pi\nu(x)\gamma_c\mbox{\bf  L
}_{2n}\mbox{\bf  Q}_{2n}\right]
\end{equation}
In (\ref{c4}) $\mbox{\bf  Q}_{2n}=U_{2n}^{-1}\mbox{\bf  L}_{\, 2n} U_{2n}$ and the
integration is over the coset space $U(2n)/U(n)\!\otimes\!\! U(n)$.

Thus, the problem reduces to evaluation of an integral over a
coset space. This type of integrals is known in the literature under
the name of zero-dimensional nonlinear $\sigma-$models and
the corresponding calculation is a standard one and is outlined in \cite{FK}.
The result is given by:
\begin{eqnarray}\nonumber
\fl {C}^{s}_{{\gamma}}( {\{\tilde z\}}) \propto \frac{\det
{\gamma}^n }{|\Delta({  \{\tilde z\}} )|^2}
\int\limits_{-1}^{1}d\lambda_1\ldots\int\limits_{-1}^{1}d\lambda_n\
\det \left[
     e^{-i\pi\nu(x)\tilde{z}_j\lambda_k}
     \right]
\det \left[
     e^{i\pi\nu(x) \tilde{z}_j^*\lambda_k}
     \right]
\prod_{j=1}^n \prod_{c=1}^M [g_c+\pi\nu(x)\lambda]
\\ \lo   \propto \frac{n!\det  {\gamma}^n}{|\Delta({  \{\tilde z\}}
)|^2}\det \!\! \left[\int\limits_{-1}^{1} \! d\lambda \prod_{c=1}^M [g_c+\pi\nu(x)\lambda]
e^{i\pi\nu(x)\lambda(\tilde{z}_j-\tilde{z}_k^*)}\right]
\end{eqnarray}

Combining this with Eqs.\ (\ref{interm}) and (\ref{CCC}),
denoting $u=\lambda\pi\nu(x)$ and restoring
the normalization we
finally see that the correlation functions have the following simple structure:
\begin{equation}\label{RRfin}
\frac{1}{N^{2n}}R_n\left(x+\frac{\tilde{z}_1}{N},
\ldots ,x+\frac{\tilde{z}_n}{N} \right)=\det\left[ K (\tilde{z}_j,\tilde{z}_k^*)\right]|_{j,k=1,\dots ,n}\ ,
\end{equation}
where the kernel $ K(\tilde{z}_j,\tilde{z}_k^*)$
is given by
\begin{eqnarray}\label{kerfin}
K(\tilde{z}_1,\tilde{z}_2^*) &=& \frac{1}{\pi}
F^{1/2}(\tilde{z}_1)F^{1/2}(\tilde{z}_2)
\int_{-\pi\nu}^{\pi\nu}  du
\prod_{c=1}^M [g_c+u]
e^{-iu (\tilde{z}_1-\tilde{z}_2^*)}
\end{eqnarray}
with $F(\tilde{z})= \sum_{k=1}^M\frac{e^{-2\im \tilde{z}g_k}} {\prod_{s(\ne
k)}(g_s-g_k)}\theta(\mbox{Im}\tilde{z})$. Introducing a "characteristic function"
\begin{equation}\label{fgam}
f_{\Gamma}(u)=\sum_{c=1}^M\ln{\left(1+\frac{u}{g_c}\right)}
\end{equation}
 and noticing that
\begin{equation}\label{rep}
\fl F(\tilde{z})=\theta(\mbox{Im}\tilde{z})\sum_{k=1}^M\frac{e^{-2\im \tilde{z}g_k}} {\prod_{s(\ne
k)}(g_s-g_k)}\equiv \left(\prod_c \frac{1}{g_c}\right)
\frac{1}{4\pi}\int_{-\infty}^{\infty} dk
\exp(-ik\im \tilde{z}-f_{\Gamma}(-ik/2))
\end{equation}
we can rewrite the kernel Eq.(\ref{kerfin})
in an equivalent form. Denoting $\tilde{z}_1=iy_1-\omega/2;\,\, \tilde{z}_2=iy_2+\omega/2$ we find that the kernel is given by :
\begin{eqnarray}\label{kerfin1}
\fl \tilde{K}(Z_1,Z_2^*)=N^2K(\tilde{z}_1,\tilde{z}_2^*)=
\frac{N^2}{4\pi^2}\int_{-\pi\nu}^{\pi\nu}du\ 
e^{-(y_1+y_2)u+ i\omega u+f_{\Gamma}(u)}
\\ \nonumber \times
\left(\int_{-\infty}^{\infty} dk_1
\exp\{-ik_1y_1-f_{\Gamma}(-ik_1/2)\}\int_{-\infty}^{\infty} dk_2
\exp\{-ik_2y_2-f_{\Gamma}(-ik_2/2)\}\right)^{1/2}
\end{eqnarray}

Using the derived expression we are going to demonstrate
 that for a system with many open channels $M\gg g$
 the statistics of resonances eventually becomes Ginibre-like. This fact coordinates with existing numerical simulations \cite{reso,numr,MS}.

To this end we consider the case of equivalent channels:
$g_1=g_2=...=g_M\equiv g$ and evaluate all the three integrals
in Eq.(\ref{kerfin1}) by the saddle-point method in the limit $M\gg g\sim 1$.
A straightforward  calculation gives
$$
\int_{-\infty}^{\infty} dk_1
\exp\{-ik_1y_p-f_{\Gamma}(-ik_p/2)\}\approx
M^{1/2-M}e^M\sqrt{2\pi}\frac{1}{y_p}(2gy_p)^Me^{-2gy_p}
$$
for $p=1,2$ whereas the integral over $u$ dominated by
the saddle-point $u_s=-g+M/y_s\,;\, y_s=y_1+y_2-i\omega$ is estimated in the limit $M\to\infty$ as
$$
M^{M+1/2}(gy_s)^{-M} e^{-M+gy_s}\sqrt{2\pi}\frac{1}{y_s}
$$
as long as $|\re u_s|\le \pi\nu$. In the opposite case $|\re u_s| > \pi\nu$ the integral is zero. Combining these expressions
and taking the absolute value of the kernel results in:
\begin{equation}\label{kerr}
|\tilde{K}(Z_1,Z_2^*)|=N^2\frac{2^{M-1}M}{\pi}
\frac{(y_1y_2)^{(M-1)/2}}{[(y_1+y_2)^2+\omega^2]^{(M+1)/2}}\quad;\quad
M\gg g
\end{equation}
This expression immediately shows us that the mean density of
resonances is$$
\rho(Z)=|\tilde{K}(Z,Z^*)|=N^2\frac{M}{4\pi y^2}=\frac{M}{4\pi (\im Z)^2}\quad;
\quad \frac{M}{2(g+\pi\nu)}\le y\le \frac{M}{2(g-\pi\nu)}$$
and zero otherwise. Such a  "cloud of resonances with a gap"
 is in agreement with the earlier result for
the mean density obtained by Haake et al.\cite{Sok} in the limit$ M\to \infty\,;\, N\to \infty$ in such a way that $M/N=fixed$.

Let us now suppose that both $Z_1$ and $Z_2$ are within the "cloud".
Then it is possible to rewrite Eq.(\ref{kerr}) as:
$$
|\tilde{K}(Z_1,Z_2^*)|=\left(\rho(Z_1)\rho(Z_2)\right)^{1/2}\left(1-
\frac{(y_1-y_2)^2+\omega^2}{(y_1+y_2)^2+\omega^2}\right)^{\frac{M+1}{2}}
$$
Taking into account that actually $y_{1,2}\ge M/(2g+2\pi\nu)\gg 1$
we see that always $y_1+y_2\gg \omega\sim 1$ whereas the relation between
$y_1-y_2$ and $\omega$ can be arbitrary. Introducing$$
\rho\left(\frac{Z_1+Z_2}{2}\right)\equiv \frac{MN^2}{\pi(y_1+y_2)^2}
$$
we see that
\begin{eqnarray}
\fl |\tilde{K}(Z_1,Z_2^*)|=\left(\rho(Z_1)\rho(Z_2)\right)^{1/2}
\left(1-\frac{\pi\rho\left(\frac{Z_1+Z_2}{2}\right)}{MN^2}
[(y_1-y_2)^2+\omega^2]\right)^{\frac{M+1}{2}}\\ \lo \nonumber \to
\left(\rho(Z_1)\rho(Z_2)\right)^{1/2}
e^{-\frac{1}{2}\pi\rho\left(\frac{Z_1+Z_2}{2}\right)|Z_1-Z_2|^2}
\end{eqnarray}
in the limit $M\to \infty$. It is further evident, that the correlations
vanish very rapidly when the distance $|Z_1-Z_2|$
between eigenvalues exceeds $M^{-1/2}$. Thus, we can safely
put $\rho(Z_1)\approx
\rho(Z_2)\approx\rho\left(\frac{Z_1+Z_2}{2}\right)$ when $M\to
\infty$ arriving finally to the Ginibre-like expression for
the kernel:
\begin{equation}\label{Ginibres}|\tilde{K}(Z_1,Z_2^*)|=\rho(Z)
\exp{-\frac{1}{2}\pi\rho\left(Z\right)|Z_1-Z_2|^2}
\end{equation}
In contrast to the Ginibre case, here the density $\rho(Z)$
inside the cloud is not constant but is rather position-dependent. It is natural to conjecture that expression Eq.(\ref{Ginibres}) constitutes the most universal form of the eigenvalue statistics
for the case of strongly non-Hermitian random matrices. This conjecture
is also supported with existing numerics, e.g. for a QCD-inspired non-Hermitian model \cite{QCD2}.

Having at our disposal the correlation functions, we can easily
repeat the calculations done in the previous sections and find
such statistical characteristics as spectral form-factor and the
number variance \cite{FTS,FM}. In particular, for the case of equivalent channels
$g_c=g, \,c=1,...,M$ the variance of the number of resonances in a strip
$0<\mbox{Re}Z<L=L_x\Delta;\, -\infty<\im Z<\infty$ is given by:
\begin{equation}\label{varr}
\fl \Sigma_2(L_x)=L_x-\frac{1}{\pi^2}\int_0^1 \frac{dk}{k^2}
\sin^2{(\pi k L_x)}\int_{-(1-k)}^{(1-k)}dv
\left[1-\frac{k^2}{(g+v)^{2}}\right]^M
\end{equation}
where we have used $\pi\nu(0)=1$. Let us discuss for simplicity
its typical features for the simplest case of a single open channel
$M=1$. We are actually interested in deviations of the number
variance from its value known for Hermitian GUE matrices.
One finds:
\begin{equation}\label{varr1}
\fl \delta\Sigma_2(L_x)=\Sigma_2(L_x)-\Sigma_2^{GUE}(L_x)
=\frac{2}{\pi^2}\int_0^1dk\sin^2{(\pi k L_x)}
\frac{1-k}{g^2-(1-k)^2}
\end{equation}
As usual, we are interested in the behaviour of the number variance for
$L_x\gg 1$, where $\Sigma_2^{GUE}(L_x)$ grows logarithmically as:
$\Sigma_2^{GUE}(L_x)\approx \frac{1}{\pi^2}\ln{L_x}+const+...$
For any $g>1$ we find that the difference
$\delta\Sigma(L_x)$ tends to a constant value $\frac{1}{2\pi^2}
\ln{[g^2/(g^2-1)]}$. This fact means that the curves
$\Sigma_2^{GUE}(L_x)$ in the asymptotic regime $L_x\gg 1$ are just shifted
upwards by a constant amount. The shift is the larger the more "open"
 is the system, i.e. the closer is the value of $g$ to unity.
The so-called "perfect coupling" case $g=1$ is known to be in many
respects specific \cite{FSR}. Physically, it describes the situation
when  direct scattering is completely absent. On a level of
number variance such a specificity is reflected in changing the
coefficient in front of the leading logarithmic growth from the GUE value
$1/\pi^2$ to a new value $3/(2\pi^2)$.

Qualitatively, the same picture holds for $M>1$: for a fixed value
of the strip widths $L_x$ the number variance is the larger, the larger is $M$ and
the closer are the coupling constants $g$ to unity.
This fact is just a
 fingerprint of a gradual "decorrelation" of resonance positions along the
real axis. As to the small-distance behaviour of the nearest-neighbour
distance distribution $p(Z_0,S\ll \Delta)$ in the complex plane
, the leading term for $S\to 0$ turns out to be always
cubic: $p(Z_0,S\ll \Delta)\propto S^3$, as long as the system is open,
in agreement with the existing numerical data \cite{numr,Main,reso}.
However, for few open channels and $g\gg 1$ the interval of cubic
behaviour turns out to be parametrically small: $S\ll g^{-1}\Delta$.
For the observation points $Z_0$ taken at regions with
asymptotically low density of resonances
 a crossover to an anomalous $S^{5/2}$ behaviour
might be observable, cf. \cite{FKS3}.

In the end of the present section we discuss briefly
an intimate relationship which actually exists between the kernel
Eq.(\ref{kerfin}) derived for a finite-rank perturbation and its Gaussian
counterpart, Eq.(\ref{kern}). For this we notice, that
 the kernel Eq.(\ref{kerfin}) derived under an assumption
of {\it positivity} of each $\gamma_c$ ( hence $g_c$), but in fact its
form Eq.(\ref{kerfin1}) describes the statistics of complex
eigenvalues for {\it arbitrary} finite-rank perturbations,
irrespective of the sign of a particular eigenvalue $\gamma_c$.
This fact can be proven by a straightforward modification of
the technique outlined in the present section. Moreover, a little thought
makes it natural to
expect that its validity is not restricted even by {\it finite-rank}
perturbations, but rather comprises the whole class of
{\it almost-Hermitian} matrices, characterized by
$N\mbox{Tr} \Gamma^2\propto \mbox{tr} H^2$ in the limit $N\to\infty$.
As a confirmation of this conjecture, let us look from this point
of view on a typical {\it Gaussian} deformation. For that case a
typical eigenvalue of $\Gamma$ is $\gamma_c\sim N^{-1/2}\ll 1$
and therefore $g_c^{-1}\approx 2\gamma_c,\,\, c=1,...,N$.
This fact allows us to expand the "characteristic function" in
Eq.(\ref{fgam}) as:
\begin{equation}\label{expansion}
f_{\Gamma}(u)|_{N\to \infty}\approx 2u\mbox{tr}\Gamma-2u^2\mbox{Tr}
\Gamma^2+....
\end{equation}
It is an easy task to show that for a Gaussian $\Gamma$ we have
typically $\mbox{Tr} \Gamma\sim N^{-1/2}\ll \mbox{Tr} \Gamma^2\sim O(1)$.
Therefore one can put safely: $f_{\Gamma}(u)\approx -2u^2\mbox{Tr} \Gamma^2$ which renders
the $k-$integrals in Eq.(\ref{kerfin1}) to be Gaussian ones.
Identifying: $\mbox{Tr}\Gamma^2=\alpha^2/4$ and
performing the Gaussian integrals
 we see that  the resulting kernel coincides with that
given in Eq.(\ref{kern}), up to an overall phase factor
$e^{ix/2(y_2-y_1)}$ which in any way does not
play any role in calculating the correlation functions of eigenvalues.

\subsection{Density of complex eigenvalues via supersymmetry approach}

Unfortunately, it is not known at the moment how to extend
the methods described above to the case of
non-Hermitian deformations of real symmetric matrices.
We consider this issue as one of the most challenging problems for
future investigations.
At the same time the mean density of complex eigenvalues
turns out to be accessible via a technique known as
Efetov supersymmetry approach and is shortly discussed below.
In fact, the method provides access to non-Hermitian deformations
of matrices interpolating between real symmetric and complex Hermitian,
as discussed in \cite{FKS3}.

As before we decompose any $N\times N$ matrix $\hat{J}$ into a sum of its Hermitian and skew-Hermitian parts and
consider an ensemble of random $N\times N$ complex matrices
$\hat{J}=\hat{H}_1+iv\hat{H}_2$
where $\hat{H}_p;\,\, p=1,2$ are both Hermitian:
$\hat{H}^{\dagger}_p=\hat{H}_p$.
The parameter $v$ is used to control the degree of non-Hermiticity.
In turn, complex Hermitian matrices $\hat{H}_p$
can always be represented as $\hat{H}_1=\hat{S}_1+iu\hat{A}_1$ and
$\hat{H}_2=\hat{S}_2+iw\hat{A}_2$, where $\hat{S}_p=\hat{S}_p^T$
is a real symmetric matrix, and $\hat{A}_p=-\hat{A}_p^T$
is a real antisymmetric one. From this point of view the parameters
$u,w$ control the degree of being non-symmetric.

In the first part of this subsection we consider the matrices $\hat{S}_1,\hat{S}_2,\hat{A}_1,
\hat{A}_2$ to be mutually statistically independent, with i.i.d. entries
normalized in such a way that
\begin{equation}\label{norm}
\lim_{N\to\infty}\frac{1}{N}{\mbox Tr}\hat{S}_p^2=\lim_{N\to\infty}\frac{1}{N}{\mbox Tr}\hat{A}_p\hat{A}_p^T=1
\end{equation}
 Such a normalization ensures that for any fixed value of the parameter $u>0$
statistics
of real eigenvalues of the Hermitian matrix
$\hat{H}=\hat{S}+iu\hat{A}$ in the limit $N\to \infty$ is identical (up to a trivial rescaling) to
that of $u=1$, the latter case being standard GUE. On the other hand, for $u\equiv 0$ real eigenvalues of the real symmetric matrix $\hat{S}$
follow a different pattern of the Gaussian Orthogonal Ensemble (GOE).
The non-trivial crossover between GUE and GOE types of statistical
behaviour happens on a scale $u\propto N^{-1/2}$ \cite{RMT1}.

Similar arguments show \cite{FKS1,FKS3}, that the
most interesting behaviour
of {\it complex} eigenvalues of non-Hermitian matrices
should be expected for the parameter $v$ being scaled in
a similar way: $v\propto N^{-1/2}$.
It is just the regime of {\it weak non-Hermiticity} which we are interested in. Under these conditions a non-Hermitian matrix $\hat J$ still
"remembers" the statistics of its Hermitian part $\hat{H}_1$.
As will be clear afterwards, the parameter $w$
should be kept of the order of unity in order to influence the statistics
of the complex eigenvalues.

Correspondingly, we scale the parameters as
\footnote{In the Letter \cite{FKS3} there is a misprint in the
definition of the parameter $\alpha$.}:
\begin{equation}\label{scale}
v=\frac{\alpha}{2\sqrt{N}};\quad u=\frac{\phi}{2\sqrt{N}}
\end{equation}
and consider $\alpha,\phi,w$ fixed of the order O(1) when $N\to \infty$.

For calculating the mean density of complex eigenvalues
$Z_k=X_k+iY_k, \quad k=1,2,...,N$
defined as
\begin{equation}\label{defden}
\rho(Z)=\sum_{k=1}^N\delta^{(2)}(Z-Z_k)=\sum_{k=1}^N
\delta(X-X_k)\delta(Y-Y_k)\equiv \rho(X,Y)
\end{equation}
one relates it to the "charge potential" \cite{ellipse}:
$$
\Phi(X,Y,\kappa)=\overline{\frac{1}{2\pi}
\ln{Det[(Z-{\cal H}_{eff})(Z-{\cal H}_{eff})^{\dagger}+\kappa^2]}}
$$
in view of the relation: $\rho(X,Y)=
\lim_{\kappa\to 0}\partial^2 \,\Phi(X,Y,\kappa)$,
where $\partial^2$ stands for the two-dimensional Laplacian,
see appendix A of \cite{FSR}.
Technically, it is convenient to introduce
 the generating function (cf.\cite{FSR})
\begin{equation}\label{genf}
{ \cal Z}=
\frac{{\mbox Det}\left[(Z-{\cal H}_{eff})(Z-{\cal H}_{eff})^{\dagger}+\kappa^2\right]}
{{\mbox Det}\left[(Z_b-{\cal H}_{eff})(Z_b-{\cal H}_{eff})^{\dagger}+\kappa^2\right]}
\end{equation}
in terms of which
\begin{equation}
\rho(Z)=-\frac{1}{\pi}\lim_{\kappa\to 0}\frac{\partial}{\partial Z^*}
\lim_{Z_b\to Z}\frac{\partial}{\partial Z_b}{\cal Z}
\end{equation}
Assuming the matrix elements of all involved matrices
 to be statistically independent, up to symmetry constraints, one
performs the ensemble averaging by the
standard trick of representing the ratio of the two
determinants in Eq.(\ref{genf}) in terms of Gaussian
superintegrals. After a set of manipulations exposed in \cite{FKS3}
one arrives at the following expression:
\begin{equation}\label{unires}
\fl \langle\rho(X,y)\rangle=\frac{N\nu(X)}{16}
\int d\mu(\hat{Q})\ \mbox{Str}\left(\hat{\sigma}_\tau^{(F)}\hat{Q}\right)
\mbox{Str}\left(\hat{\sigma}_\tau\hat{Q}\right)\exp{-S(\hat{Q})}
\end{equation}
\begin{equation}
\fl S(\hat{Q})=-\frac{i}{2}y\mbox{Str}\left(\hat{\sigma}_\tau\hat{Q}\right)-
\frac{a^2}{16}\mbox{Str}\left(\hat{\sigma}_\tau\hat{Q}\right)^2+
\frac{b^2}{16}\mbox{Str}\left(\hat{\tau}_2\hat{Q}\right)^2-\frac{c^2}{16}
\mbox{Str}\left(\hat{\sigma}\hat{Q}\right)^2
\end{equation}
where we introduced the scaled imaginary parts $y=\pi\nu(X)NY$,
with $\nu(X)$ being the mean (semicircular) eigenvalue density.
We also used the notations: $a^2=\left(\pi\nu(X)\alpha\right)^2,\quad b^2=\left(\pi\nu(X)\phi\right)^2,\quad c^2=\left(\pi\nu(X)\alpha w\right)^2$.
Here the integration goes over the set of $8\times 8$
supermatrices $\hat{Q}$ satisfying the constraint $\hat{Q}^2=-1$.
The symbols $\mbox{Str,\,Sdet}$ stand here for the graded
trace and the graded determinant, correspondingly.

Properties of these matrices and the integration measure
$d\mu(\hat{Q})$ can be found in \cite{Efbook}. The matrices
$\hat{\sigma}_{\tau}^{(F)}\,,\, \hat{\sigma}_{\tau}$ and $\hat{\tau}_2$
in Eq.(\ref{unires}) are some fixed block-diagonal $8\times 8$ supermatrices,
and can be found in \cite{FKS3}.
The whole expression is just a general $\sigma-$ model
representation of the mean density of complex eigenvalues in the
regime of weak non-Hermiticity.
We expect it to be universal and applicable even beyond the case of matrices with independent entries, after replacement of the semicircular density $\nu(X)$ with actual density of real eigenvalues of the Hermitian part. The expression is parametrised by 
$a,b$ and $c$. The   parameters controll the degree of non-Hermiticity
($a$), and symmetry properties of the
Hermitian ($b$) and non-Hermitian ($c$) parts of the ensemble.

Still, in order to obtain an explicit expression for the
density of complex eigenvalues one has to
evaluate the integral over the set of supermatrices
$\hat{Q}$. In general, it is an elaborate
task due to the complexity of that manifold. At the present moment such an evaluation was successfully performed
for three "pure" cases: those of
almost-Hermitian matrices \cite{FKS1}, real almost-symmetric
matrices \cite{Efnonh} and complex
symmetric matrices \cite{SFT}.
The first case (which is technically the simplest one)
corresponds to $\phi\to\infty$, that is $b\to \infty$.
Under this condition only that part of the matrix
$\hat{Q}$ which commutes with
$\hat{\tau}_2$ provides a nonvanishing contribution
and the resulting integral \cite{FKS1} proves to be equivalent to the "diagonal" part $K(Z,Z^*)$ of the kernel Eq.(\ref{kern}), as expected.

The second nontrivial case for
which the result is known explicitly is due to Efetov \cite{Efnonh}.
This is the case of real matrices with "weak asymmetry".
In the present notations this case corresponds to
the limit
$\phi\to 0; w\to \infty$ in such a way that the product
$\phi w=\tilde{c}$ is kept fixed.
The density of complex eigenvalues turns out to be given by:
\begin{eqnarray}\label{Efres}
\fl \rho_X(y)=\delta(y)\int_0^1 dt\exp{(-\tilde{c}^2t^2/2)}\\ \nonumber +
2\sqrt{\frac{2} {\pi}}\frac{|y|}{\tilde{c}}
\int_1^{\infty}du \exp \left(-\frac{2y^2u^2}{\tilde{c}^2} \right)
\int_0^1  dt t \sinh (2t|y|) \exp{(-\tilde{c}^2t^2/2)},
\end{eqnarray}

The first term in this expression shows that everywhere in the regime of
"weak asymmetry" $\tilde{c}<\infty$ a finite fraction of
eigenvalues remains on the real axis. Such a behaviour is qualitatively
different from that typical for the
case of "weak non-Hermiticity" $\tilde{a}<\infty$, where eigenvalues
acquire a nonzero imaginary part with probability one.
In the limit $\tilde{c}>>1$ the portion of real eigenvalues
behaves like $\tilde{c}^{-1}$. Remembering the normalization
of the parameter $v$, Eq.(\ref{norm}), it is easy to see that
for the case of $v=O(1)$ the number of real eigenvalues should scale
as $\sqrt{N}$.
 The fact that of the order of $N^{1/2}$ eigenvalues of strongly asymmetric
real matrices stay real was first found numerically by
Sommers et al. \cite{ellipse,Lehm1}, and proven by Edelman
\cite{Edel}.

At last, the case of complex symmetric matrices ($b=c=0$ in the
present notation)
turns out to be most involved technically. The result of evaluating
the integral Eq.(\ref{unires}) over the coset space \cite{SFT}
turns out to be:
\begin{equation}\label{comsym}
\fl \left\langle\rho_X(y)\right\rangle=-\frac{1}{16\pi}
\frac{\partial}{\partial y}\int_{-1}^1d\lambda F^{-2}(\lambda)
\int_{-\infty}^{\infty}d\lambda_1 F(i\lambda_1)
\int_{\lambda_1}^{\infty}d\lambda_2 F(i\lambda_2)
\frac{(\lambda_2-\lambda_1)(2\lambda-i\lambda_1-i\lambda_2)}
{(\lambda-i\lambda_1)^2(\lambda-i\lambda_2)}
\end{equation}
where
\begin{equation}\label{f}
F(\lambda)=e^{\lambda y + a^2\lambda^2}\frac{1}{(1-\lambda^2)^{1/2}}\ .
\end{equation}

Actually, the integrals over $\lambda_1,\lambda_2$ as they stand
in Eq.(\ref{comsym})  should be understood
in the sense of a principal value. To be precise, the correct
expression is equal to the half-sum of two integrals with integration
contours encircling the singular point
$\lambda_1=-i\lambda; \lambda_2=-i\lambda$ from above and from below.

Of course, the
supersymmetry approach is not restricted to anti-Hermitian deformations $i\hat{\Gamma}$ whose
entries are statistically independent as assumed above,
but can be used as well for fixed finite-rank deformations. This allows one
to calculate the density of resonance poles in the complex energy
plane. In fact, such a calculation was the first application of the supersymmetry approach to non-Hermitian
problems \cite{FS} and paved the way to all later
developments in \cite{FKS1,Efnonh,QCD,SFT}.

One finds the following $\sigma-$model representation
for the distribution of scaled
resonance widths $y=\frac{\pi\im Z}{\Delta}<0$ (measured in units of the local
mean level spacing $\Delta$ of the closed system) for
the resonances whose positions are within
a narrow window around the central point  $X=0$ of the spectrum. 
\begin{eqnarray}\label{widres}
\fl \langle\rho_X(y)\rangle=\frac{1}{16}\int d\mu(\hat{Q})\ 
\mbox{Str}\left(\hat{\sigma}_\tau^{(F)}\hat{Q}\right)\mbox{Str}\left(\hat{\sigma}_\tau\hat{Q}\right)
\exp{\frac{i}{2}y\mbox{Str}\left(\hat{\sigma}_\tau\hat{Q}\right)}
\\ \nonumber \times
\prod_{a=1}^M\mbox{ Sdet}^{-1/4}\left[1-\frac{i}{2g_a}
\left\{\hat{Q},\hat{\sigma}_{\tau}\right\}\right]\ .
\end{eqnarray}
Here $\left\{\hat{Q},\hat{\sigma}_{\tau}\right\}=
\hat{Q}\hat{\sigma}_\tau+\hat{\sigma}_\tau\hat{Q}$ stands
for the anticommutator.

 Performing the integration
over the manifold of the supermatrices $\hat Q$ (which is of different
form for different symmetry classes) one finds for the case of broken invariance \cite{FS} the distribution
coinciding with that following from the kernel Eq.(\ref{kerfin}),
whereas for the case of preserved invariance the distribution is given by
Eq.(\ref{comsym}), but with the function $F(\lambda)$
replaced with the following expression \cite{SFT}:
\begin{equation}\label{f1}
F_{\Gamma}(\lambda)=
\frac{1}{\prod_{c=1}^M(g_c-\lambda)^{1/2}}
\frac{e^{\lambda y}}{(1-\lambda^2)^{1/2}}
\end{equation}
Such a replacement looks completely natural from the point of view of
the correspondence pointed out in the end of preceding section, see Eq.(\ref{expansion}).

Finally,  we would like to mention a finite-rank
generalization of Efetov's result, Eq.(\ref{Efres}) as performed
in \cite{FTS}. Namely, we consider the ensemble of real asymmetric
matrices of the form: $H+A$, with $H$ being a real symmetric (taken
from GOE) and $A$ being a {\it fixed} real
antisymmetric: $A_{ij}=-A_{ji}$ of a
finite rank $M$. As is well-known,
a general antisymmetric matrix $A$ of even dimension
can be brought by an orthogonal
rotation to the following block-diagonal form: $\diag(A_1,....,A_N)$,
with each block $A_i$ being $2\times 2$ matrix of the form
$A_i=\left(\begin{array}{cc} 0& \mu_i \\ -\mu_i& 0\end{array}
\right)$. Invoking the argument of the rotational
invariance of GOE matrices, it is enough to consider deformations
to be of such block-diagonal form. We restrict our attention to the
finite-rank case, when all $\mu_i=0$ for $i>M$. Introducing the
scaled variable $y=\pi\nu(X)N \im Z$ we find for the rescaled density
$\rho_X(y)=\langle \rho(Z)\rangle [N\pi\nu(X)]^{-2}$
the following expression \cite{FTS}:
\begin{eqnarray}\label{Efres1}
\fl \rho_X(y)=\delta(y)\int_0^1 dt
e^{\frac{1}{2}[f_A(\pi\nu t)+f_A(-\pi\nu t)]}+
\frac{1} {2\pi}
\int_0^{1}dt t \sinh(|y|t)
e^{\frac{1}{2}[f_A(\pi\nu t)+f_A(-\pi\nu t)]}
\\ \lo \nonumber \times \int_{|y|}^{\infty}ds
\int_{-\infty}^{\infty} dk \exp\{-iks-\frac{1}{2}
[f_A(i\pi\nu k)+f_A(-i\pi\nu k)]\}
\end{eqnarray}
 where the function $f_A(z)$ is given by the same expression as
Eq.(\ref{fgam}), but with $\gamma_c$ replaced by $\mu_i$, i.e. $g_c$ replaced by $(\mu_i +1/\mu_i)/2$.

Our interest in the problem was stimulated by the
papers \cite{QCD} claiming that expression Eq.(\ref{Efres}), 
derived for Gaussian $A$,
describes well the complex eigenvalue density even for non-random
antisymmetric deformations by matrices
$A=\left(\begin{array}{cc} 0& \mu \hat{{\bf 1}}\\ -\mu \hat{\bf{1}}& 0\end{array}
\right)$, with a constant $\mu$ being of the order of $\mu\sim
N^{-1/2}$. This fact immediately follows
from Eq.(\ref{Efres1}). Indeed, for
$M\propto N$ and the typical
$\mu_i$ of the order of $N^{-1/2}$ the function $f_{A}(v)$
can be expanded up to a first non-vanishing order such that
$f_A(v)+f_A(-v)\propto v^2\mbox{Tr} A^2$, cf. Eq.(\ref{expansion}). The corresponding expression then indeed coincides with Eq.(\ref{Efres}).

\section{Eigenvalue Statistics of Subunitary Matrices}

\subsection{Truncations of random unitary matrices}

Let us first consider truncated matrices drawn from the 
circular unitary ensemble (CUE).
Let $U=\left(\matrix{A & B \cr C & D \cr}\right)$
be an $ N \times N $ matrix from CUE and $A$ a subunitary $(N-M) \times (N-M)$ matrix. To make contact to the general case discussed in the introduction we note that putting $B=0$ and $D = 0$ corresponds to a matrix  
$\hat{A}=\hat{u} \sqrt{{\bf 1}-\hat T}$ with $T_i =0$ for $i=1,...,N-M$ and$T_i=1$ for $i=N-M+1,...,N $. This matrix has the same nonzero eigenvalues as the subblock $A$ of the matrix $U$ and we determine below the joint density of those eigenvalues inside the complex unit circle.

The joint density of elements of $U$ can be written as
\begin{eqnarray}
\fl P(U) \propto \delta\left(A^{\dagger}A+C^{\dagger}C-{\bf 1}\right)\delta\left(A^{\dagger}B+C^{\dagger}D\right)
\delta(B^{\dagger}B+D^{\dagger}D-{\bf 1})
\label{G1}
\end{eqnarray}
with appropriate matrix $\delta-$functions. Integrating out $B$ and $D$
we obtain as joint density of elements of $A$
\begin{equation}
 P(A) \propto  \int dC\  \delta (A^{\dagger}A+C^{\dagger}C-{\bf 1})
\label{G2}
\end{equation}
with a $2M(N-M)$ dimensional integration over the complex parameters $C$.
Performing these remaining integrations is possible after representing
the $\delta$-function in Eq.(\ref{G2}) as a Fourier
transform over the set of Hermitian matrices ${\cal F}$
of the size $(N-M)\times (N-M)$. Then integrals over $C$ are
essentially Gaussian and can be immediately performed yielding:
\begin{equation}\label{trunc2}
 P(A) \propto  \int d{\cal F}\  \det{{\cal F}}^{-M}
\exp\{i\mbox{\small Tr} {\cal F} ({\bf 1}-A^{\dagger}A)\}
\end{equation} 
The matrix integrals of this type emerge frequently in various
context, and most
 recently were calculated in \cite{IS1,IS2}. 
For the case $M\ge N/2$ the probability $P(A)$ is given
by a simple, non-singular expression:
\begin{equation}\label{trunc3}
 P(A) \propto \det\{{\bf 1}-A^{\dagger}A\}^{2M-N}
\theta\{{\bf 1}-A^{\dagger}A\}
\end{equation}  
where the matrix $\theta-$function is unity when its argument 
is a positive definite matrix and is zero otherwise.

In the opposite, most interesting case $M<N/2$
the integral still can be calculated explicitly
\cite{IS2}  but the result is complicated and contains a product
of many singular ($\delta-$functional) contributions. 

We are however mostly interested in finding the joint probability
density of complex eigenvalues rather than just distribution of
the matrix $A$ itself. This can be achieved directly from
the equation Eq.(\ref{G2}) if we again use the Schur 
decomposition $A =V (z+ \hat{R})V^{-1}$ extensively exploited by us
in Sec.2. Here the
transformation $V$ is unitary, $z$ is a diagonal
matrix of the complex eigenvalues of $A$ and $\hat{R}$ 
is strictly upper triangular. A procedure allowing one to integrate
out matrices $V$, the $(N-M)(N-M-1)/2$ complex parameters
$R_{ij}$ and the $M(N-M)$ complex parameters $C$ is described in
some detail in \cite{KS}. The resulting expression is however 
rather simple:
\begin{equation}
P(\{ z\} ) \propto \prod_{i<j}^{1... N-M}\vert z_i-z_j \vert^2 \prod_{i=1}^{N-M} (1-\vert
z_i\vert^{ 2})^{M-1}\theta(1-\vert z_i\vert^{ 2})\
\label{G6}
\end{equation}
This distribution is very analogous to the Ginibre  ensemble and
all correlation functions can be found readily by the method of orthogonal
polynomials. In fact, the powers $z^{n-1}$ 
are already orthogonal inside the unit circle of the complex plane. 
An equivalent method is to
consider the joint density $P(\{ z \})$ as the 
absolute square of a Slater determinant of normalized wave functions
$$ \phi_n(z) = z^{n-1} w(\vert z\vert^2)/\sqrt{N_n}$$, with 
$N_n$ standing for a normalization factor and the domain of
integration being $|z|\le 1$.
The kernel, which determines all correlation 
functions is given by $$K(z_1,z_2^*) = \sum_{n=1}^{N-M} (z_1z_2^*)^{n-1} w(\vert z_1\vert^2)
w(\vert z_2\vert^2) /N_n.$$ 
Further defining $x=r^2=\vert z\vert^2$
we can, for example, easily write down the averaged density
of complex eigenvalues at point $|z|$ as 
\begin{equation}
 P(r)={2r\over {N-M}}{(1-x)^{ M-1}\over( M-1)!}
\left({d\over dx}\right)^{M}
{(1-x^N)\over 1-x}.
\label{G8}
\end{equation}

There are two important limiting cases for large $N$:
either $\mu=M/N$ fixed or
$M$ fixed. For fixed $\mu$ and $N \to\infty$ we find the
scaling behaviour:
\begin{equation}
 P(r)=     {\mu \over {1- \mu} }   { 2r \over(1-r^2)^2}
\label{prmu}
\end{equation}
 for $r^2<{1 -\mu}$ and
$P(r)=0$ otherwise. The distribution shows a gap near the unit circle
and qualitatively resembles one obtained for resonances in the chaotic scattering
problem for large number of channels, see Haake et al.\cite{Sok}. 
In this - "strongly nonunitary" -
limit one can also simplify the cluster function given
by $ Y(z_1,z_2)= \vert K(z_1,z_2^*)\vert^2$:
\begin{equation}
Y(z,z+\delta) = (N-M)^2 \rho (z)^2\ \exp (-\pi (N-M)\rho (z)\ |\delta|^2)
\label{cluster}
\end{equation}
This is just the Ginibre behaviour Eq.(\ref{Ginibres}) with the
 distance $\delta$ rescaled by the local mean level distance 
$1/\sqrt{(N-M)\rho(z)}$ given by (\ref{prmu}) through $\rho(z)=P(r)/2\pi r$. The same can be shown for the nearest neighbour distance
distribution obtained by Grobe et al. \cite{diss} and
applied to a damped chaotic kicked top.

In the other limit of fixed $M$ and $N$ to $\infty$ the matrices may be
considered
as weakly nonunitary, and we recover exactly the universal resonance-width
distribution \cite{FS,FSR} for perfect coupling to $M$ channels with
$y=N(1-r)$
\begin{equation}
 \rho(y)={y^{M-1}\over(M-1)!}\left({-d\over dy}\right)^{M}{1-
e^{-2y}\over 2y}.
\label{weak}
\end{equation}
Similarly the cluster function obtained in this limit can be shown to coincide
with the one obtained in the previous section for chaotic
scattering with a finite number $M$ of perfectly coupled channels.
The statistics (\ref{weak}) has also been found by Kottos and Smilansky
\cite{graphs} for chaotic scattering on graphs and by Gl\"uck et al.\cite{KolRev} for a model of crystal electron
in the presence of dc and ac fields.
In both of these works the $S$-matrix is reduced to the resolvent of a
subunitary matrix as is investigated in the present subsection.

\subsection{General finite-rank deviations from unitarity}
In what follows we assume all eigenvalues of the deviation
matrix $\hat{T}=1-\hat{G}$ are smaller than unity: $T_i<1$,
 but the resulting expressions turn out to be valid 
in the limiting cases some $T_i=1$ as well.

We again start with the Schur decomposition $\hat{A}=\hat{U}(\hat{Z}+\hat{R})\hat{U}^{\dagger}$ of the matrix $\hat A$ in terms of a unitary matrix $\hat{U}$, a diagonal matrix of the eigenvalues $\hat{Z}=\{z\}$
and a upper triangular matrix $\hat{R}$. One can satisfy oneself, that
the eigenvalues $z_1,...,z_N$ are generically not degenerate, provided
all $T_i<1$. Then, the measure written in terms of new variables is given by $d\hat{A}=|\Delta(\{z\})|^2 d\hat{R}d\hat{Z}d\mu(\hat U)$, where
the first factor is just the squared Vandermonde determinant of eigenvalues $z_i$
and $d\mu(\hat U)$ is the invariant measure on the unitary group.
The joint probability density of complex eigenvalues is then given
by: 
%%%%%%%%%%%%%%%%%%%%%%%%%%%%%%%%%
\begin{eqnarray}\label{9}
\fl {\cal P}(\{z\})\propto |\Delta(\{z\})|^2  \int d\mu(\hat 
U)d\hat{R}\
\delta\left((\hat{Z}+\hat{R})(\hat{Z}+\hat{R})^{\dagger}-
\hat{U}^{\dagger}\hat{G}\hat{U}\right)\ .
\end{eqnarray}
The integration over $\hat{R}$ can again be   performed using its  
triangularity (similar to the case of "truncated" matrices
considered above). We give here a derivation in more detail. 
 Let us consider the columns of the $N\times N$ unitary matrices $U$ as
$N$-component (mutually orthogonal) vectors $\,{\bf a}_l,\
l=1,...,N$, so that  $\hat U =({\bf a}_1,{\bf  
a}_2,....,{\bf a}_N)$. 
Then the $\delta$-functions in Eq.(\ref{9}) imply 
\begin{eqnarray}\label{diagonal}
|z_i|^2 + \sum_{k(<i)}|R_{ki}|^2 +{\bf a}_i^{\dagger}\hat T {\bf a}_i 
=1
\end{eqnarray}
and for $i<j$:
\begin{eqnarray}\label{offdiagonal}
z_i^{*}R_{ij} +\sum_{k(<i)}R_{ki}^{*}R_{kj} +{\bf a}_i^{\dagger}\hat T 
{\bf a}_j=0
\end{eqnarray}
 Integration over the complex parameters $R_{ij}$ can easily be done 
and yields the Jacobian ${1 / \prod_{i<j}|z_i|^2}$. 
 The remaining integration over  
the unitary group should be done with the product of the diagonal 
$\delta$-functions as integrand. 
There the solution $R_{ij}$ of Eq.(\ref{offdiagonal}) has to 
be inserted. We can find that solution employing 
the hierarchical structure of the  
equations. It can easily be checked, that the solution is given by
\begin{eqnarray}\label{solution}
R_{ki}=-{1\over z_k^{*}}{\bf a}_k^{\dagger}{1\over{1-\hat T({\bf 
a}_1\otimes  
{\bf a}_1^{\dagger}+...+{\bf a}_{k-1}\otimes {\bf 
a}_{k-1}^{\dagger})}}\hat T {\bf a}_i.
\end{eqnarray}
One may also show that Eq.(\ref{offdiagonal}) can be extended to $i=j$ 
such that
\begin{eqnarray}\label{Rii} \nonumber
-z_i^{*}R_{ii} = \sum_{k<i}|R_{ki}|^2 +  {\bf a}_i^{\dagger} \hat T 
{\bf a}_i \\ = {\bf a}_i^{\dagger}{1\over{1-\hat T({\bf a}_1\otimes 
{\bf a}_1^{\dagger}+...+{\bf a}_{i-1}\otimes {\bf a}_{i-1}^{\dagger})}}\hat 
T {\bf a}_i\ .
\end{eqnarray}
This implies
\begin{eqnarray}\label{P1}
\fl {\cal P}(\{z\})\propto |\Delta(\{z\})|^2{1\over 
\prod_{i<j}|z_i|^2}\\
\nonumber
\times \int d\mu(\hat U)\prod_i \delta (|z_i|^2 -1 +
{\bf a}_i^{\dagger}{1\over{1-\hat T({\bf a}_1\otimes 
{\bf a}_1^{\dagger}+...+
{\bf a}_{i-1}\otimes {\bf a}_{i-1}^{\dagger})}}\hat T{\bf a}_i)\ .
\end{eqnarray}
Finally we use the identity
\begin{eqnarray}\label{trace} \nonumber
\fl 1-{\bf a}_i^{\dagger}{1\over{1-\hat T({\bf a}_1\otimes 
{\bf a}_1^{\dagger}+...+{\bf a}_{i-1}\otimes {\bf
a}_{i-1}^{\dagger})}}\hat T {\bf a}_i 
={\det (1-\hat T({\bf a}_1\otimes {\bf a}_1^{\dagger}+...+{\bf  
a}_{i}\otimes {\bf a}_{i}^{\dagger}))\over \det (1-\hat T({\bf 
a}_1\otimes  
{\bf a}_1^{\dagger}+...+{\bf a}_{i-1}\otimes {\bf a}_{i-1}^{\dagger}))}
\end{eqnarray}
%%%%%%%%%%%%%%%%%%%%%%%%%%%%%%%%%%%
and introducing the projector matrices $\hat{P}_l=\mbox{diag}
\left(1,...1,0,...,0\right)$ after a simple algebra we arrive at
the following expression: 
\begin{eqnarray}\label{110}
\fl {\cal P}(\{z\})\propto |\Delta(\{z\})|^2
\int d\mu(\hat U) \prod_{l=1}^N
\delta\left(|z_1|^2...|z_l|^2-
\det\left[{\bf 1}-\hat{T}\hat{U}\hat{P}_l\hat{U}^{\dagger}\right]\right)\ .
\end{eqnarray}
The remaining integration over the unitary group
can be done, in essence, similar to
the corresponding procedure for non-Hermitian matrices, see
Section 2.2. Actual calculations are, however, more involved and we outline the main steps of the full solution below
extending \cite{FSJETP}.

First, introduce $N\times N$  matrices of rank $l$:
$\hat{Q}_l=\sum_{i=1}^l{\bf a}_i\otimes {\bf a}^{\dagger}_i$ ,
so that $\hat{U}\hat{P}_l \hat{U}^{\dagger}=\hat{Q}_l$.
Due to the fact that only $M$ out of $N$ eigenvalues
of the matrix $\hat{T}$ are non-zero, both the matrices $\hat{Q}_l$ and
the vectors ${\bf a}$ can be effectively taken to be of the
size $M$ (it amounts to changing the unspecified normalization
constant in Eq.(\ref{110})). We then redefine the matrix $\hat{T}$
as $\hat{T}=\mbox{diag}(T_1,...,T_M)=\hat{\tau}^{\dagger}\hat{\tau}$,
i.e. the matrix of transmission coefficients  $0 \le T_a \le 1$.

Writing down the corresponding constraints in a form of
$\delta-$ functions, one can represent the
expression Eq.(\ref{9}) in the form:
\begin{eqnarray}\label{P({z})a} \nonumber
{\cal P}(\{z\}) &\propto&  |\Delta(\{z\})|^2
\int \prod_{i=1}^Nd^2{\bf a_i} \ \delta({\bf 1}-\sum_{j=1}^N{\bf a_j} \otimes {\bf a_j^{\dagger}}) \\
  &\times& \prod_{l=1}^N
\delta\left(|z_1|^2...|z_l|^2-
\det\left[{\bf 1}-\hat{T}\sum_{j=1}^l{\bf a_j} \otimes {\bf a_j^{\dagger}}\right]\right)\ .
\end{eqnarray}
Here ${\bf a_i}, \ i=1,...,N $ are $M$-dimensional complex vectors building a $N \times M$ subblock of a unitary matrix $U$ and we define $d^2{\bf a_i} := \prod_{j=1}^M \re d{\bf a_i}^{(j)}  \im  d{\bf a_i}^{(j)}$. The unitary constraint is forced by the inserted $\delta$-function. Obviously $|z_i|^2
= { \det({\bf 1}-\hat T \sum_{j=1}^i {\bf a_j} \otimes {\bf a_j^{\dagger}})/\det({\bf 1}-\hat T \sum_{j=1}^{i-1}{\bf a_j} \otimes {\bf a_j^{\dagger}})} \le 1\ $ as it should be for complex eigenvalues of a contraction.

We want to integrate over the unitary group which amounts to integrate over all complex vectors ${\bf a_i}$.
 We increase again the number of integration variables  introducing as independent variables    matrices ${q_i}$ through the dyadic product $ q_j ={\bf a_j} \otimes {\bf a_j^{\dagger}}$. The $M\times M$ Hermitian matrix $q_i$  contains   in general more independent real variables than the complex vector ${\bf a_i}$. Therefore we insert the identities $1= \int dq_i\ \delta(q_i- {\bf a_i} \otimes {\bf a_i^{\dagger}})$ with integrations over the Hermitian matrices $q_i$ and the corresponding $\delta$-functions for independent elements. Then
(\ref{P({z})a}) takes the form
\begin{eqnarray}\label{P({z})q} \nonumber
{\cal P}(\{z\}) &\propto&  |\Delta(\{z\})|^2
\int \prod_{i=1}^N dq_i\ \delta({\bf 1}-\sum_{j=1}^N q_j ) \prod_{i=1}^N \int d^2{\bf  a} \  \delta(q_i-{\bf a}\otimes {\bf a^{\dagger}})\\
  &\times& \prod_{l=1}^N
\delta\left(|z_1|^2...|z_l|^2-
\det\left[{\bf 1}-\hat{T}\sum_{j=1}^l q_j\right]\right)\ .
\end{eqnarray}
Finally, we introduce $\hat Q_l = \sum_{i=1}^l q_i$ and observe that Equ.( \ref{P({z})q}) aquires the form
\begin{eqnarray}\label{111}
\fl {\cal P}(\{z\})\propto |\Delta(\{z\})|^2
\int \left(\prod_{l=1}^{N-1} d\hat{Q}_l\right) \prod_{l=1}^N
\delta\left(|z_1|^2...|z_l|^2-
\det({\bf 1}-\hat{T}\hat{Q}_l)\right)
\\ \lo \nonumber \times
\prod_{i=1}^N\int d^2{\bf a}\  \delta\left(\hat{Q}_i-\hat{Q}_{i-1}-
{\bf a}\otimes{\bf a}^{\dagger}\right)\ ,
\end{eqnarray}
where the matrices $\hat{Q}_l$ are considered to be unconstrained
$N\times N$ Hermitian. We also used the orthonormality condition
$\hat{Q}_N={\bf 1} $ as well as the convention $\hat{Q}_0={\bf 0} $.

\subsection{Rank-1 deviations from unitarity}
The remaining integration over the vectors ${\bf a_i}$ still requires a quite lengthy calculation. The general case will be considered in the next section. A particular case which can be solved relatively easy is the case
of a rank-one deviation from Hermiticity: $M=1$, corresponding to a
system with only one open channel. This example is quite instructive and the exposition below follows \cite{prel}.
Evaluating the integrals Eq.( \ref{111}) quite straightforwardly  we
arrive at a very simple expression:
\begin{equation}\label{11}
{\cal P}(\{z\})\propto T^{1-N}|\Delta(\{z\})|^2
\delta\left(1-T -|z_1|^2...|z_N|^2\right)\ .
\end{equation}
provided $0\le |z_l|\le 1$ for all eigenvalues, and zero otherwise.
Here $0<T<1$ is the only non-zero eigenvalue of the deviation matrix.

 Eq.(\ref{11}) can be used to extract all n-point correlation functions defined as in Eq.~( \ref{corfunR}).
To achieve this it is convenient to use the Mellin transform with respect to the variable $\zeta = 1-T$:
\begin{equation}\label{12}
\tilde{R}_n(s;\{z\}_n)=\int_{0}^{\infty}d\zeta\  \zeta^{s-1}
\left[(1-\zeta)^{N-1}R_n(\{z\}_n)\right]\ .
\end{equation}
It is easy to notice that such a transform brings ${\cal P}(\{z\})$ to the form
suitable for exploitation of the orthogonal polynomial method
\cite{Mehta}. The corresponding polynomials   $p_k(z)=z^k \sqrt{k+s}$ are orthonormal with respect to the weight $f(z)=|z|^{2(s-1)}/\pi$ inside the unit circle $|z|\le 1$.
Following the standard route we find:
\begin{equation}\label{13}
\tilde{R}_n(s;\{z\}_n)\propto
\frac{\det{\left[K(z_i,z_j^*)\right]}|_{(i,j)=1,...,n}}{s(s+1)...(s+N-1)}
\end{equation}
where the kernel is
\begin{eqnarray}\label{13a}
K(z_1,z_2^*)&=&(f(z_1)f(z_2))^{1/2}\sum_{k=0}^{N-1}p_k(z_1)p_k(z_2^*)\\
\nonumber & = &
{|x|^{s-1} \over \pi }\left( s\phi(x)+x\frac{d}{dx}\phi(x)\right)|_{x=z_1z_2^*}
\end{eqnarray}
and $\phi(x)=(x^N-1)/(x-1)$.
Thus, the expression  Eq.(\ref{13}) can be rewritten as:
\begin{eqnarray}\nonumber
\tilde{R}_n(s;\{z\}_n)&\propto& \frac{\prod_{l=1}^n|z_l|^{2(s-1)}}{s(s+1)...(s+N-1)}
\sum_{l=0}^ns^l q_l(\{z\}_n);\\
\nonumber
q_0(\{z\}_n)&=&\det{\left[x\frac{d}{d
x}\phi(x)|_{x=(z_i z_j^*)}\right]}|_{i,j=1,...,n}\\ \nonumber &\ldots&\\
\nonumber q_n(\{z\}_n)&=&\det{\left[\phi(x)|_{x=(z_i z_j^*)}\right]}|_{i,j=1,...,n}
\end{eqnarray}
and can easily be   Mellin-inverted yielding finally the original correlation functions in the following form:
\begin{eqnarray}
\fl R_n(\{z\}_n )|_{|z_n|\le 1}\propto
T^{1-N}\theta(T-1+a)\sum_{l=0}^n q_l(\{z\}_n)
 \left(\frac{d}{da}a\right)^l  \left.
\left[\frac{1}{a}\left(1-\frac{1-T}{a}\right)\right]^{N-1}
\right|_{a=\prod_{i=1}^n|z_i|^2}
\end{eqnarray}
where $\theta(x)=1$ for $x\ge 0$ and zero otherwise.
This equation is exact for arbitrary $N$. Let us now investigate the limit
$N\gg n\ge l$ and use, to the leading order:
\[\left(\frac{d}{da}a\right)^l
\left[\frac{1}{a}\left(1-\frac{\xi}{a}\right)\right]^{N-1}\approx
\left(\frac{N\xi}{a-\xi}\right)^l\frac{1}{a}\left(1-\frac{\xi}{a}\right)^{N-1}
\quad,\quad \xi=1-T\,.
\]
This allows one to rewrite the correlation function as a determinant:
\begin{eqnarray}\label{20}
R_n(\{z\}_n )&\propto&
T^{1-N}\frac{1}{a}\left(1-\frac{1-T}{a}\right)^{N-1}
\theta(T-1+a) \\ \nonumber
&\times & \det_{i,j=1,...,n}{\left(\frac{N(1-T)}{a-1+T}\phi(x)
+x\frac{d}{d x}\phi(x)\right)|_{x=z_i z_j^*}}\ .
\end{eqnarray}

Further simplifications occur after taking into account that
the eigenvalues $z_i$ are expected to concentrate
typically at distances  of order of $1/N$ from the unit circle.
Introduce new variables
$y_i,\theta_i$ according to $z_i=(1-y_i/N) e^{i\theta_i}$
and consider $y_i$ to be of the order of unity when $N\to \infty$.
As to the phases $\theta_i$  their typical separation scales as:
$\theta_i-\theta_j=O(1/N)$.
 Now it is straightforward to perform the limit
$N\to \infty$ explicitly and bring Eq.( \ref{20}) to the final form:
\begin{eqnarray}\nonumber
\fl R_n(\{z\}_n )&\propto&e^{-g\sum_{i=1}^ny_i}\det{\left[\int_{-1}^{1}d\lambda (\lambda+g)
e^{-\frac{i}{2}\lambda\left(N(\theta_i-\theta_j)-i(y_i+y_j)\right)}\right]_{i,j=1,\dots,n}}
\end{eqnarray}
with $g=2/T-1$.\footnote{The coupling constants $g_a$ in this section are rescaled and correspond to $g_a/\pi \nu (X)$ of subsection {\it 2.2}.There is no difference at $X=0$. The coupling constants are always $\ge 1$.} 

The expression above coincides in every detail with that for  rank-one deviations from Hermiticity, see Eqs.(\ref{RRfin},\ref{kerfin}) provided one fixes before the rescaled coupling constant $g$ and then remembers that the mean linear
density of phases $\theta_i$ along the unit circle is $\nu=1/(2\pi)$.
The non-unitary matrices provide therefore an alternative way
to address analytically "local-in-spectrum" universal features
 arising in the theory
of chaotic scattering. Sometimes to deal with subunitary matrices is
technically more advantageous than with their non-Hermitian
counterparts. As an illustration of this statement
we use subunitary matrices for
 extracting the distribution $f(\Gamma)$ of the width $\Gamma=2\mbox{Im}z$ of the most narrow resonance among $n=W/\Delta>>1$ falling
in a narrow window $[E-W/2,E+W/2]$ in the vicinity
of a given point $E$ in the spectrum. The latter quantity is of
great interest in the theory of random lasing \cite{Schom}
since it is related to fluctuations of the lasing threshold.
The functional form of the distribution was
found in those papers by employing plausible qualitative arguments
of virtual statistical independence of the widths of the neighbouring
resonances. For the simplest single-channel
case we are able to show analytically that \cite{FM}
\begin{equation}\label{res3}f(\Gamma)= \frac{\pi g n}{\Delta}e^{- \pi g n\Gamma/\Delta  }
\end{equation}
The general case can be treated very similar.
The distribution Eq.( \ref{res3}) is exactly that suggested in \cite{Schom}, but with renormalized effective
coupling $g=\frac{1}{2\pi\nu}(\gamma+\gamma^{-1})$ replacing the combination $\gamma/2\pi{\nu}$. We see that the two expressions coincide only within the weak coupling limit $\gamma\to 0$, whereas the difference amounts to the factor $2$ in the exponent for the case
of perfect coupling $\gamma=1$.
Let us briefly comment on the way of deriving the distribution
Eq.( \ref{res3}).
Instead of extracting such a quantity from the joint probability density Eq.( \ref{jpd}) we find it technically easier to consider its counterpart Eq.( \ref{11}) and interprete the parameter
$N$ in Eq.(\ref{11}) as the number $n$ of resonances in the window. We know that
in the limit $n>>1$ the eigenvalues $z_k=r_ke^{\theta_k}$ are situated in a narrow vicinity of the unit circle
and their statistics is indistinguishable from that of the complex eigenenergies ${\cal E}$,
when the latter are considered {\it locally}, i.e. on the distances comparable with the mean spacing $\Delta$. In particular, the distances $1-r_k$ from the unit circle should be
interpreted as the widths of the resonances.
To calculate the
distribution Eq.( \ref{res3}) we first notice that the form of
Eq.( \ref{11}) allows one to integrate out the phases $\theta_i$
by noticing that:
\begin{eqnarray}
&&\int_0^{2\pi}\frac{d\theta_1}{2\pi}\ldots
\int_0^{2\pi}\frac{d\theta_n}{2\pi}
\prod_{k<j}^{1...n}|r_k e^{i\theta_k}-r_j e^{i\theta_j}|^2
=\mbox{Per}(r^2_1,...,r^2_n)
\end{eqnarray}where we denoted $\mbox{Per}(x_1,...,x_n)=
\sum_{\{\alpha\}}x_1^{\alpha_1}\ldots x_n^{\alpha_n}$,
and the summation  goes over all possible permutations $\{\alpha\}=(\alpha_1,...,\alpha_n)$ of the set
$1,...,n$ (in fact here we deal with an object known as "permanent", hence the notation). In this way we come to the joint
probability density of the radial coordinates only. Such a density
written in terms of the variables $R_i=r^2_i$ has the following form\begin{eqnarray}\label{radial}
{\cal P}_T(R_1,...,R_n)\propto T^{1-N}\mbox{Per}(R_1,...,R_n)
\delta\left(1-T-R_1\ldots R_n\right)\ .
\end{eqnarray}
Identifying resonance widths with the distances $1-r_i$, the probability distribution of the
resonance $r_1$ closest to the unit circle: $r_1>r_2,r_3,...,r_n$ is naturally described in terms of the function:
\begin{eqnarray}\label{funct}
p_T(R_1)=\int_0^{R_1}dR_2\ldots \int_0^{R_1}dR_n {\cal P}_T(R_1,...,R_n)\ .
\end{eqnarray}
Again it is convenient to multiply this expression with $T^{N-1}$ and
to perform the Mellin transform with respect to $1-T$. Then the
integration over $R-$variables is simple to perform and after inverting the
Mellin transform one finds:
\begin{eqnarray}\label{radial1}
p_T(R_1)=\theta\left(R^n_1-1+T\right)\frac{1}{T^{n-1}}\frac{d}{d R_1}
\left[R_1^{n/2}\left(1-\frac{1-T}{R_1^n}\right) \right]^{n-1}\ .
\end{eqnarray}
The distribution Eq.(\ref{res3}) follows from Eq.(\ref{radial1}) after rescaling $R_1=1-2y/n$,
$y=\pi\Gamma/\Delta$,
performing the limiting procedure $n\to \infty$ and remembering $\Delta=2\pi/n$.

\subsection{Rank $M>1$ deviations from unitarity}

Coming back to the general case we find it convenient to change: $\hat{T}^{1/2}\hat{Q}_l\hat{T}^{1/2}\to
\hat{Q}_l$ and separate integration over eigenvalues and eigenvectors of
matrices $\hat{Q}_l$. The latter can be performed in a recursive way
$l\to l+1$, with the multiple use of the
Itzykson-Zuber-Harish-Chandra Eq.(\ref{IZHC}) formula.
After quite an elaborate manipulation, demonstrated in detail in appendix A2  one finally arrives at the
following representation:
\begin{eqnarray}\label{main}
\fl {\cal P}(\{z\})\propto \frac{\det^{M-N}(\hat{T})}{\det({\bf 1}-\hat{T})
\prod_{c_1<c_2}\left(T_{c_1}-T_{c_2}\right)}
\prod_{c_1<c_2}\left(\frac{\partial}{\partial \tau_{c_1}}
-\frac{\partial}{\partial \tau_{c_2}}\right) \\ \lo\nonumber
\times \int d\hat{\lambda}
e^{-i\mbox{Tr}{\hat{\tau}\hat{\lambda}}}|\Delta(\{z\})|^2
\prod_{k=1}^Nf(\ln{|z_k|^2},\hat{\lambda}),
\end{eqnarray}
where we defined the diagonal matrices of size $M$ as:
$\hat{\tau}=\mbox{diag}(\tau_1,...,\tau_M)\,,\,
\hat{\lambda}=\mbox{diag}(\lambda_1,...,\lambda_M)$
and used the notations: $\tau_c=\ln{(1-T_c)}$ and
\begin{equation}
f(a,\hat{\lambda})=i^{M-1}\sum_{q=1}^M\frac{e^{i\lambda_qa}}
{\prod_{s(\ne q)}(\lambda_q-\lambda_s)}\quad.
\end{equation}

The distribution Eq.(\ref{main}) is written for $|z_k|^2\le 1$ for
any $k=1,...,N$ and vanishes otherwise. It shows great similarity to
the pair of equations (\ref{P(Z1)},\ref{start}) describing eigenvalues of matrices deviating from Hermiticity.
The remarkable feature of the distribution Eq.(\ref{main}) is that it allows for calculation of all
$n-$point correlation functions for arbitary $N,n,M$ with help of
 the method of orthogonal polynomials. Again, the particular case
 $M=1$ \cite{prel} is quite instructive and can be recommended to follow
first for understanding of the general formulae outlined below.

To this end, we write
\begin{eqnarray}
\fl |\Delta(\{z\})|^2
\prod_{k=1}^N f(\ln{|z_k|^2},\hat{\lambda})=
\prod_{k=1}^N N_k(\hat{\lambda})\det{\left[\sum_{n=1}^N
\frac{(z_iz_j^*)^{n-1}}{N_n(\hat{\lambda})}
f(\ln{|z_j|^2},\hat{\lambda})\right]_{i,j=1,...N}}\quad.
\end{eqnarray}
where the constants $N_n(\hat{\lambda})$ are provided by the orthonormality
condition:
\begin{equation}
\int_{|z|^2\le 1}d^2z z^{m-1}(z^*)^{n-1}f(\ln{|z|^2},\hat{\lambda})
=\delta_{m,n}N_n(\hat{\lambda})\quad,
\end{equation}
which yields after a simple calculation $N_n(\hat{\lambda})=\pi\prod_{c=1}^M
\frac{1}{(n+i\lambda_c)}$.

Now, by applying the standard machinery of orthogonal polynomials
\cite{Mehta} one can find the correlation function:
\begin{eqnarray}
R_n( z_1,...,z_n)=\frac{N!}{(N-n)!}\int d^2z_{n+1}...d^2z_N{\cal
P}(\{z\}) \quad
\end{eqnarray}
as given by:
\begin{eqnarray}\label{genres}
\fl R_n( z_1,...,z_n)\propto \hat{{\cal D}}
\int d\hat{\lambda}
e^{-i\mbox{Tr}{\hat{\tau}\hat{\lambda}}}
\prod_{k=1}^N N_k(\hat{\lambda})
\det{\left[K(z_i,z_j^*;\hat{\lambda})\right]}_{(i,j)=1,...,n}
\end{eqnarray}
where the kernel $K$ is defined as:
\begin{equation}
K(z_1,z_2^*;\hat{\lambda})=\frac{1}{\pi}
\sum_{n=1}^N\det{(i\hat{\lambda}+n{\bf 1})}(z_1z_2^*)^{n-1}
f(\ln{|z_2|^2},\hat{\lambda})
\end{equation}
and the differential operator $\hat{\cal D}$ is just the expression
in front of the $\lambda-$ integral in Eq.(\ref{main}).
It is worth noting that we used a version of the kernel which
is not symmetric with respect to its arguments $z_1,z_2$.
This choice is legitimate, though unconventional, and is
dictated mainly by our wish to avoid square roots which may induce
complications when working with complex-valued expressions.

In principle, all $\lambda-$ integrations in
Eq.(\ref{genres}) can be performed explicitly and the resulting
 formulae provide the desired general solution of the problem.
However, for arbitary $N,M,n$ the results obtained in that way
are still quite cumbersome. Some examples are given
in \cite{FSJETP}. In the theory of quantum chaotic scattering we, however,
expect a kind of universality in the semiclassical limit. Translated
to the random matrix language such a limit corresponds
to $N\to \infty$ at fixed $n,M$.  Still, extracting the asymptotic behaviour
of the correlation function $R_n(z_1,...,z_n)$ from Eq.(\ref{genres})
in such a limit is not a completely straightforward task.
A useful trick is to notice that Eq.(\ref{genres}) can be rewritten
as:
\begin{eqnarray}\label{gr1}
\fl R_k( z_1,...,z_n)\propto
\sum_{q_1=1,...,q_n=1}^M F_{q_1,...,q_n}\left(\{T_c;z\}\right)\quad,\\
\fl F_{q_1,...,q_n}\left(\{T_c;z\}\right) = {\cal B}\{T_c\}
\det{\left[\sum_{k=1}^N\left(\frac{-\partial}{\partial \tau_1}+k\right)
... \left(\frac{-\partial}{\partial
\tau_M}+k\right)(z_iz_j^*)^{k-1}\right]}_{i,j=1,...,n}\\
\fl \nonumber \times \prod_{c_1< c_2}^{1...M}
\left(\frac{\partial}{\partial\tau_{c_1}}
-\frac{\partial}{\partial \tau_{c_2}}\right)
 \int \prod_{c=1}^M\left(d\lambda_c
\frac{\exp{-i\lambda_c\tau_c}}{\prod_{l=1}^N(l+i\lambda_c)}\right)
\frac{e^{-i\sum_{j=1}^n\lambda_{q_j}\ln|z_j|^2}}
{\prod_{j=1}^n\prod_{s(\ne q_j)}^{1...M}(\lambda_{q_j}-\lambda_s)}  \quad,
\end{eqnarray}
where we used the notation:
\[ {\cal B}\{T_c\}=
\frac{1}{\prod_{c_1<c_2}\left(T_{c_1}-T_{c_2}\right)}
\prod_{c=1}^M\frac{T_c^{M-N}}{(1-T_c)}\quad.
\]
Introducing now the auxiliary differential operator
$\hat{{\cal D}}_{q_1,...,q_n}=\prod_{j=1}^n\prod_{s(\ne q_j)}^{1...M}
\left(\frac{\partial}{\partial \tau_{q_j}}-
\frac{\partial}{\partial \tau_s}\right)$
and considering its action upon the ratio $F_{q_1,...,q_n}/
{\cal B}\{T_c\}$ one can satisfy oneself that in the limit $N\gg M,n$
the leading contribution to $F_{q_1,...,q_n}$ is given by:
\begin{eqnarray}\label{interm1}
\fl F_{q_1,...,q_n}\propto \prod_{c=1}^M\theta(1-\tilde{T}_c)
\frac{(1-\tilde{T}_c)}{(1-T_c)}
\left(\frac{\tilde{T}_c}{T_c}\right)^{N-M}
\prod_{j=1}^n\prod_{s(\ne q_j)}^{1...M}
\left(\frac{1}{ T_{q_j}}-
\frac{1}{T_s}\right)^{-1}
\det{\left[K(z_i,z_j^*;\{\tilde{T_c}\})\right]_{(i,j)=1,...,n}}
\end{eqnarray}
where the kernel is given by
\begin{eqnarray}\label{kernel}
K(z_i,z_j^*;\{\tilde{T_c}\})&=&
\sum_{k=1}^N\prod_{c=1}^M\left[(N-M)\frac{1-\tilde{T}_c}
{\tilde{T}_c}+k-1\right](z_iz_j^*)^{k-1}\\
\nonumber &=& \prod_{c=1}^M\left[(N-M)\frac{1-\tilde{T}_c}
{\tilde{T}_c}+x\frac{d}{dx}\right]\frac{1-x^N}{1-x}\left.\right|_{x=z_i z_j^*}\end{eqnarray}
and we used the notation:
$\tilde{T}_c=1-\exp{\left(\tau_c-
\sum_{j=1}^n\delta_{q_j,c}\ln{|z_j|^2}\right)}$.

Further simplifications occur after taking into account that the
eigenvalues $z_i$ are, in fact,  concentrated
typically at distances  of order of $1/N$ from the unit circle.
Then it is natural to introduce new variables
$y_i,\phi_i$ according to $z_i=(1-y_i/N) e^{i\theta_i}$
and consider $y_i$ to be of the order of unity when $N\to \infty$.
First of all, one immediately finds that:
\begin{equation}\label{rat}
\lim_{N\to \infty}
\prod_{c=1}^M\left(\frac{\tilde{T}_c}{T_c}\right)^{N-M}
=\exp{\left(-2\sum_{j=1}^n y_j\frac{1-T_{q_{j}}}{T_{q_{j}}}\right)}
\end{equation}
 As to the phases $\theta_i$, we expect their typical separation scaling as: $\theta_i-\theta_j=O(1/N)$.
 Now it is straightforward to perform explicitly the limit
$N\to \infty$ in Eq.(\ref{kernel}). Combining all factors together, one
brings the correlation function
Eq.(\ref{gr1}) to the final form
 coinciding in every detail with that obtained
 for random GUE matrices deformed by a finite rank 
anti-Hermitian
perturbation, see Eqs.(\ref{RRfin},\ref{kerfin}), with $g_c=2/T_c-1$.
One should only remember that the mean
density of phases $\theta_i$ along the unit circle is $\nu=1/(2\pi)$
and take this factor into account
 replacing the (semicircular) density factor $\nu(x)$ whenever necessary after fixing the rescaled coupling constants $g_a$.
This completes the proof of equivalence, in the large N limit,
 of spectral properties for finite-rank deviations from Hermiticity and from unitarity.

We have seen already that the correlation function of moments of characteristic polynomials played a very important role
in the theory of non-Hermitian matrices. It is natural to consider
similar objects for general subunitary matrices $\hat{A}$ as well. So far
only the simplest objects of this kind were
calculated \cite{prel1}:
\begin{equation}\label{21}
I(z_1,z_2)=\left\langle\left[\det{\left(z_1{\bf 1}-\hat{A}\right)}
\det{\left(z_2^*{\bf 1}-\hat{A}^{\dagger}\right)}\right]^n\right\rangle_A
\end{equation}
where the angular brackets stand for the averaging over the
probability density in Eq.(\ref{0}). We present the
final result, which has the form of a Hankel determinant,
 in terms of eigenvalues $G_i=1-T_i$ of $\hat{G}$:
\begin{eqnarray}\label{22}
\fl I(z_1,z_2)=\prod_{j=0}^{n-1}\frac{(N+n+j)!}{j!(j+1)!(N+j)!}
\det{\left[f_{i+j}\right]}_{i,j=0,...,n-1}
\end{eqnarray}
\begin{eqnarray*}
\fl f_{i+j}=\sum_{k=0}^N (z_1z_2^*)^{N-k}\frac{(N+i+j-k)!(2n-2+k-i-j)!}
{(N+2n-1)!}
\sum_{1\le i_1<i_2<...<i_k\le N} G_{i_1}G_{i_2}...G_{i_k}\ .
\end{eqnarray*}
The particular case $n=1$ of the above formula was presented
in \cite{prel}.
In fact, the expression Eq.(\ref{22}) is valid for any $G_i>0$, being not
at all restricted to the subunitary case $0<G_i<1$.
For random unitary matrices all $G_i=1$ and one can show that
the expression above are indeed 
in agreement with the known result for the moments of
characteristic polynomilas of unitary matrices \cite{sec,sec1}.

\section{Statistics of Non-Orthogonal Eigenvectors}

Let us denote $\left.|R_k\right\rangle$ and $\left\langle L_k|\right.$
to be the right and the left eigenvectors of the matrix ${\cal H}$
corresponding to the eigenvalue $z_k$. This means
\begin{eqnarray}
{\cal H}\left.|R_k\right\rangle={z}_k\left.|R_k\right\rangle
\quad,\quad \left\langle L_k|\right.{\cal H}=\left\langle
L_k\right.|{z}_k\\{\cal H}^{\dagger}\left.|L_k\right\rangle=
{z}^*_k\left.|L_k\right\rangle
\quad,\quad \left\langle R_k|\right.{\cal H}^{\dagger}=\left\langle
R_k|\right.{z}^*_k\ .
\end{eqnarray} These eigenvectors form a complete, bi-orthogonal set and can be normalized to satisfy
\begin{equation}
\left\langle L_m\right.|\left.R_n\right\rangle=\delta_{mn}
\end{equation}
The most natural way to characterize the non-orthogonality
of the eigenvectors is to consider the statistics of the overlap
 matrix ${\cal O}_{mn}=\left\langle L_m\right.|\left.L_n\right\rangle
\left\langle R_m\right.|\left.R_n\right\rangle$. This matrix naturally enters many calculations that operate
with non-Hermitian Hamiltonians,  e.g. in the description of the particle
escape from the scattering region ("norm leakage"\cite{SS}).
Moreover, the entries $O_{nm}$ have a direct physical meaning
for lasing media: diagonal ones are the so-called Petermann (or "excess noise")  factors accessible experimentally, whereas
the off-diagonal entries represent cross-correlations between the
thermal or quantum noise emitted into different eigenmodes \cite{Pet1}.

Thus, two correlation functions \cite{CM}: the {\it diagonal} one
\begin{equation}\label{dia}
O({z})=\left\langle \frac{1}{N}\sum_{n}{\cal O}_{nn}
\delta\left({z}-{z}_k\right)\right\rangle_{{\cal H}_N}
\end{equation}
and the {\it off-diagonal} one
\begin{equation}\label{ndia}
O({z},{z}')=
\left\langle \frac{1}{N}\sum_{n\ne m}{\cal O}_{nm}
\delta\left({z}-{z}_n\right)
\delta\left({z}'-{z}_m\right)\right\rangle_{{\cal H}_N}
\end{equation}
are natural characteristics of average non-orthogonality of eigenvectors
corresponding to the resonances whose positions in the complex
plane are close to the complex energy ${z}$. Here $\delta(z)$ stands for the two-dimensional $\delta$-functions of the complex
variable $z$. Let us note that for any ensemble with orthogonal eigenvectors and complex
eigenvalues $z$ (e.g. for the so-called {\it normal} matrices) one obviously has $O(z)$ equal to the mean density of
complex eigenvalues. For the same situation the off-diagonal
correlator vanishes: $O(z_1,z_2)\equiv 0$.

Both diagonal and off-diagonal eigenvector correlators were
introduced and explicitly calculated for the strongly non-Hermitian
(Ginibre) ensemble of non-Hermitian matrices by Chalker and Mehlig  \cite{CM}. For the ensemble ${\cal J}_N=H-i\hat{\Gamma}$
 pertinent to chaotic scattering both types of eigenvector correlators were found
for the regime of very strongly overlapping resonances
when widths typically much exceed the mean separation \cite{MS}. We already discussed that such a regime corresponds to
a large number $M\gg 1$ of open channels, and a kind of self-consistent Born approximation (or equivalent approximation schemes, see \cite{Janik}) is justified in this case.
A general non-perturbative expression for the diagonal correlator
$O(z)$ can be extracted from the following relation \cite{Janik}:
\begin{equation}
O(z)=\frac{1}{\pi}\lim_{\epsilon\to 0}\left\langle
\left(\mbox{Tr}\frac{\epsilon}{(z-{\cal J}_N)(z^*-{\cal
J}_N^{\dagger})+\epsilon^2 {\bf 1}} \right)^2\right\rangle\ .
\end{equation}
Although a direct calculation of the right-hand side for weakly
non-Hermitian matrices is difficult,
Frahm et al. \cite{Schom} managed to get to the result valid for any number $M$ of open channels by employing a kind of heuristic analytic continuation scheme suggested in \cite{FSR} and
 known to reproduce the exact expression for the resonance widths distribution,
like that in Eq.(\ref{weak}). Therefore we have good reasons
 to expect the result for the eigenvector correlator
found in \cite{Schom}\begin{eqnarray}\label{o1}
\fl \langle K\rangle=O(z)/\rho(z)=
1+\frac{2 {\cal F}(y)}{{\cal F}_1\left(y\right)
{\cal F}_2\left(y\right)}\quad,\quad
{\cal F}\left(y\right)=-\int_0^ydx
{\cal F}_1\left(x\right)\frac{d}{dx}{\cal F}_2\left(x\right)
\end{eqnarray}
where $y=\frac{2\pi Y}{\Delta}$ and
\begin{eqnarray}
\fl {\cal F}_1\left(y\right)=\frac{y^{M-1}}{(M-1)!}e^{-gy}\quad,\quad
{\cal F}_2\left(y\right)=(-1)^Me^{gy}\left(\frac{d}{dy}\right)^M
\left(e^{-gy}\frac{\sinh{y}}{y}\right)
\end{eqnarray}
should be valid  as well, though it is still desirable
to verify this formula independently. No non-perturbative results for the off-diagonal
eigenvalue correlator $O({z}_1,{z}_2)$
 were reported so far, to the best of our knowledge.

In a recent paper \cite{FM} exact non-perturbative expressions for both diagonal and off-diagonal eigenvector correlators were provided for
the case of a system with broken time-reversal invariance
 coupled to continuum via a single open channel. The expression for the diagonal correlator was found to agree with the formula Eq.(\ref{o1})
and was verified by comparing the analytical expression with the result of direct numerical diagonalization of the matrices ${\cal
J}_N$. In fact, numerically it is easier to compute smoothed
averages, such as the mean number of eigenvalues $\langle
n(L_x,L_y)\rangle$ inside the rectangle
$A=\left(\begin{array}{c}-L_x/2\le \re {z}\le L_x/2\quad,\quad
0\le \im {z}\le L_y \end{array}\right)$.
Such a quantity can be obtained from the mean density $\rho({z})$
 by a simple integration over the rectangular domain $A$.
Similarly one can define the function $O_1(L_x,L_y)$ as the integral
of the diagonal correlator $O({z})$ over the same domain.
Numerically the latter quantity should be compared with
$\sum_{{z}_k\in A}{\cal O}_{kk}$.

The formula for the off-diagonal correlator $O({z}_1,{z}_2)$ found in \cite{FM} is given by:
\begin{eqnarray}\label{res2}
\fl O({z}_1,{z}_2)=N
\left(\frac{\pi\nu}{\Delta}\right)^2e^{-g(y_1+y_2)}
\int_{-1}^{1}d\lambda_1\int_{-1}^{1} d\lambda_2
(g+\lambda_1)(g+\lambda_2)e^{i\omega(\lambda_1+\lambda_2)}
\\ \nonumber \times
e^{-y_2(\lambda_1-\lambda_2)/2}
\left[ e^{y_1(\lambda_1-\lambda_2)/2}-e^{-y_1(\lambda_1-\lambda_2)/2}\right]\end{eqnarray}
where we introduced $\mbox{Im}z_{1,2}=X_{1,2}=X\mp \Omega$, assumed that
$\Omega\sim \Delta$ and denoted $\omega=
\pi\Omega/\Delta$. Again, for numerical reasons it is easier
to calculate the smoothed average $\sum_{{z}_m,{z}_n,\in A}{\cal O}_{mn}$ which should be compared with the corresponding integral of the
function $O({z}_1,{z}_2)$.

Let us now outline the derivation of the formulae
Eqs.(\ref{o1},\ref{res2}). The main idea is to use the fact
that the complex eigenvalues ${z}_k$ of ${\cal H}_{eff}={\cal J}_N$
are resonances, i.e.
the poles of the $M\times M$ scattering matrix $\hat{S}({{\cal E}})$
in the complex energy plane ${\cal E}$. 
Denoting by $V$ a (nonunitary!) matrix of (right) 
 eigenvectors of ${\cal H}_{eff}=V\hat{Z}V^{-1}$
we have, in particular, the identities:
\begin{eqnarray}\label{au}
\fl \left\{V^{\dagger}
\left({\cal H}_{eff}-{\cal H}_{eff}^{\dagger}\right)V\right\}_{mn}
=(z_n-z^*_m)\left\{V^{\dagger}V\right\}_{mn}\\ \nonumber
\fl \left\{V^{-1}
\left({\cal H}_{eff}^{\dagger}-{\cal H}_{eff}\right)
\left[V^{-1}\right]^{\dagger}\right\}_{nm}
=(z^*_m-z_n)\left\{V^{-1}\left[V^{-1}\right]^{\dagger}\right\}_{nm}
\end{eqnarray}
Let us also note that according to the definition the 
entries of the overlap matrix 
${\cal O}_{mn}$ are given in terms of 
$V$ as ${\cal O}_{mn}=\left\{V^{\dagger}V\right\}_{mn}
\left\{V^{-1}\left[V^{-1}\right]^{\dagger}\right\}_{nm}$.

Substituting the relation
$E-{\cal H}_{eff}=V(E-\hat{Z})V^{-1}$
into the formula for the scattering matrix Eq.(\ref{def}) we can extract
the behaviour of the scattering matrix elements when the (complex)
energy parameter ${\cal E}$ approaches an individual pole:
\begin{equation}\label{residue1}
\fl S_{ab}\left({\cal E}\to z_n\right)\approx
-2\pi i \left(W^{\dagger}V\right)_{an}
(V^{-1}W)_{nb}\frac{1}{{\cal E}-z_n}
\end{equation}
A similar formula holds for the Hermitian conjugate of the scattering
matrix:
\begin{equation}\label{residue2}
\fl \left[S^{\dagger}\right]_{ba}\left({\cal E}\to z_m\right)
\approx 2\pi i 
\left(W^{\dagger}\left[V^{-1}\right]^{\dagger}\right)_{bm}
(V^{\dagger}W)_{ma}\frac{1}{{\cal E}^{*}-z^*_n}
\end{equation}
These two equations immediately yield:
\begin{eqnarray}\label{residue3}
\fl \mbox{Tr}\left\{S\left({\cal E}_1\to
z_n\right)\left[S^{\dagger}\right] \left({\cal E}_2\to z_m\right)
\right\}=\\ \nonumber
\fl \frac{4\pi^2}{({\cal E}_1-z_n)({\cal E}_2^{*}-z^*_n)}
\left(V^{-1}WW^{\dagger}\left[V^{-1}\right]^{\dagger}\right)_{nm}
\left(V^{\dagger}\,WW^{\dagger}\,V\right)_{nm}
\end{eqnarray}
Now we notice that $2\pi i W W^{\dagger}=
{\cal H}_{eff}-{\cal H}_{eff}^{\dagger}$ and further exploit the
relations Eq.(\ref{au}). This immediately yields the following
relation between the trace of the S-matrix residues and the 
overlap matrices of the left and right eigenvectors: 
\begin{eqnarray}\label{residues}
\fl \mbox{Tr}\left[\mbox{Res}\,\hat{S}({\cal E})
\mbox{Res}\,\hat{S}^{\dagger}(\tilde{{\cal E}}^{*})
\right]_{{\cal E}\to{z}_{n}\,\,,\,\, \tilde{{\cal E}}^{*}
\to{z}^{*}_{m}}=\left( {z}_m^*-{z}_n\right)
\left({z}_n-{z}^*_m\right)\,\,{\cal O}_{mn}\ .
\end{eqnarray}

Such a relation is of general validity, but for 
an arbitary number of open channels
$M$ it seems to be of no obvious utility, 
due to difficulties in evaluating the 
ensemble average of the trace of the residues on the
left-hand side. However for the particular case of a single open
channel $M=1$ the scattering matrix coincides with its determinant
and therefore has a different
representation as (see Eq.(\ref{dets1}))
\begin{equation}\label{1ch}
S({\cal E})=\prod_{k=1}^N\frac{{\cal E}-{z}^*_k}
{{\cal E}-{z}_k},\quad \mbox{hence} \quad
S^{\dagger}({\cal E}^*)=\prod_{k=1}^N\frac{{\cal E}^*-{z}_k}
{{\cal E}^*-{z}^*_k}\end{equation}
In fact, this formula is model-independent and
follows, up to an irrelevant "non-resonant" phase factor,
from the requirement of $S$-matrix analyticity 
in the upper half-plane and unitarity for real energies.
This expression substituted in Eq.(\ref{residues}) yields immediately
the relation:
\begin{equation}\label{overlap}
{\cal O}_{mn}=\frac{\left({z}_n-{z}^*_n\right)
\left({z}_m-{z}^*_m\right)}
{\left({z}_n-{z}^*_m\right)^2}
\prod_{k(\ne n)}^N\frac{{z}_n-{z}^*_k}{{z}_n-{z}_k} \prod_{k(\ne m)}^N\frac{{z}^*_m-{z}_k}{{z}^*_m-{z}^*_k}
\end{equation}
giving the eigenvector non-orthogonality overlap matrix
in terms of the complex eigenvalues ${z}_k$. This provides us with a
possibility to find the diagonal and off-diagonal correlators,
Eqs.(\ref{dia},\ref{ndia}), by averaging ${\cal O}_{mn}$ over the known joint
probability density of complex eigenvalues for the
single-channel scattering system Eq.(\ref{jpd}).
Indeed, one may notice that
\begin{eqnarray}\label{dia1}
\fl O({z})=
\frac{\tilde{\gamma_1}^{N-2}}{\gamma^{N-1}}
e^{-\frac{1}{2}
\left[N\gamma^2-(N-1)\tilde{\gamma_1}\right]}
e^{-\frac{N}{4}({z}^2+{z}^{*2})}
\left\langle \det[{\left({z}-{\cal J}^{\dagger}\right)
\left({z}^*-{\cal J}\right)}]\right\rangle_{\tilde{\cal J}_{N-1}}
\end{eqnarray}
where $\tilde{\cal J}_{N-1}$ stands for the non-Hermitian matrix of
the same type as ${\cal J}_N$ but of
the lesser size $(N-1)\times(N-1)$, and with coupling $\gamma$ replaced by a modified coupling $\tilde{\gamma_1}=\gamma-\mbox{Im}{z}$.
Analogously
\begin{eqnarray}\label{dia2}
\fl O({z}_1,{z}_2)=\frac{\tilde{\gamma_2}^{N-3}}{\gamma^{N-1}}
e^{-\frac{1}{2}\left[N\gamma^2-(N-2)\tilde{\gamma_2}\right]}
e^{-\frac{N}{4}\sum_{n=1}^{2}({\cal E}_n^2+{z}_n^{*2})}
({z}_1-{z}_1^*)({z}_2-{z}_2^*)\\
\lo
\nonumber \times \left\langle\det[{\left({z}_1-{\cal J}^{\dagger}\right)
\left({z}_1^*-{\cal J}^{\dagger}\right)
\left({z}_2-{\cal J}\right)
\left({z}_2^*-{\cal J}\right)
}]\right\rangle_{\tilde{\cal J}_{N-2}}
\end{eqnarray}
where $\tilde{\cal J}_{N-2}$ is of
the size $(N-2)\times(N-2)$, and with coupling $\gamma$ replaced by a modified coupling $\tilde{\gamma_2}=\gamma-\mbox{Im}{z}_1
-\mbox{Im}{z}_2$.

In this way the problem is again reduced to calculating a correlation function
of the characteristic polynomials of large non-Hermitian matrices as that in Eq.(\ref{CC_}). The scaling limit $N\gg 1$ such that $\mbox{Im}{z}_{1,2}=\Gamma_{1,2}\sim2\Omega=\mbox{Re}\left(
{z}_{1}-{z}_{2}\right)\sim \Delta\propto N^{-1}$
of the resulting expressions just yields the formulae Eqs.(\ref{o1},\ref{res2})
above.

\section{Conclusions and perspectives}

In the present paper we discussed
properties of random matrix ensembles which can be
naturally called {\it non-Hermitian/non-unitary deformations} of
classical Hermitian/unitary ensembles.

 Guided by the example of resonances in quantum chaotic scattering, we revealed the existence of a non-trivial regime of {\it weak} non-Hermiticity.
The regime can be defined as that for which the imaginary
part $\mbox{Im} Z$ of a typical complex eigenvalue is of the same order
as the mean eigenvalue separation $\Delta$ for the corresponding Hermitian
counterpart $\hat{H}$. We systematically demonstrate that for such a regime the
correlation functions of complex eigenvalues describe a non-trivial
crossover from the Wigner-Dyson statistics of real eigenvalues
typical for Hermitian/unitary random matrices to the Ginibre statistics in the
complex plane typical for ensembles with strong non-Hermiticity.
The latter is defined by the condition $<\mbox{Tr} H^2>\propto< \mbox{Tr} \Gamma^2>$ when $N\to \infty$.
Another important new feature of the non-Hermitian matrices is the emerging non-orthogonality of eigenvectors, and we discussed a few results available
in the literature.

 Among challenging problems for future research we would like to mention extending our understanding to other symmetry classes
of weakly non-Hermitian matrices, most importantly to complex symmetric and
real asymmetric cases. In fact, the joint probability densities of complex eigenvalues for some particular cases are known, see e.g.\cite{Sokr,Lehm1},
but no progress in extracting any correlation function beyond the mean
eigenvalue density was reported so far.
For non-perturbative calculations (replica trick) of the mean eigenvalue density for various non-Hermitian ensembles see the recent paper \cite{NK}.
A consistent non-perturbative treatment of non-Hermitian matrices with "chiral" symmetry is still an open problem as well.
Next to nothing is known about statistics of non-orthogonal eigenvectors for    other classes of non-Hermitian matrices, in particular, for the important case of the time-reversal invariant scattering.

Our considerations were mainly restricted to  a Gaussian probability measure for the Hermitian counterparts of the considered matrices. An interesting open issue is the universality of the obtained
results versus modification of the probability
measures for Hermitian and anti-Hermitian parts.
Relatively little is known in this direction and the available results
are restricted by some specific types of the probability measure \cite{QCD6}.
Within the non-linear $\sigma-$model formalism
one can demonstrate a kind of universality by non-rigorous, heuristic methods for matrices with independent entries. A challenging problem
is to elevate our understanding
of properties of almost-Hermitian
random matrices to the level typical for their Hermitian counterparts.

Some interesting questions still remain to be answered even for non-Hermitian deformations of GUE matrices. To ensure their applications to the theory of
chaotic scattering, we first have to improve our knowledge
on eigenvector statistics. In particular, one needs to study
the case of more than one channel and the problem of understanding
 fluctuations of the non-orthogonality overlap matrix ${\cal O}_{mn}$.

 From all these points of view a most perspective direction seems to be
the method of reducing various correlation functions of interest to products (and, in the general case, ratios!) of   spectral determinants (characteristic polynomials).
For Hermitian random matrices closely related objects
were recently quite intensively treated  by various analytical
techniques \cite{brezin1} and the issue of universality
proven to be amenable to a rigorous mathematical treatment \cite{charpol}.

Recently, there was   considerable interest in understanding interplay
between the weak non-Hermiticity and Anderson localization \cite{GNV} and in
statistics of the resonances in systems with localized eigenfunctions \cite{loc}. Clearly, this issue deserves further attention.

All questions about non-Hermitian matrices can equally be asked for
matrices deviating from unitarity. Actually, some statistical properties
of general subunitary matrices  were under investigation recently
as a model of scattering matrix for systems with absorption,
see \cite{abs1}. One may hope that this class of random matrices will find growing applications in the future.

\section{Acknowledgements}
The authors are much obliged to B.A.Khoruzhenko, B.Mehlig, M.Titov and K. {\. Z}yczkowski for
collaboration on different aspects of
non-Hermitian random matrices discussed in the present paper.
Communications and discussions with many colleagues over the years, especially with G.Akemann, Y.Alhassid, A.Garcia-Garcia, S.Grebenschikov, T.Guhr, I.Guarneri, F.Haake, G.Hackenbroich, T.Kottos, J.Keating, V.Mandelshtam, L.Pastur, D.Savin, H.Schomerus, T.Seligman, V.Sokolov, T.Wettig, H.-A.Weidenmueller, and J. Verbaarschot
on various questions related to the subject are much appreciated.

This work was supported by SFB 237 "Unordnung und grosse Fluktuationen"
and by EPSRC grant  GR/13838/01 "Random matrices close to unitary
or Hermitian" .

\appendix
\section{Various integrals over the unitary group}
\subsection{Itzykson-Zuber integral for the case of
matrices having reduced rank}

Denote ${\bf u}_c\,,\,c=1,...,N$ columns of
the unitary $N\times N$ matrix $\hat{U}\in U(N)$.
They are orthonormal $N-$component complex vectors:
${\bf u}_{c_1}^{\dagger}{\bf u}_{c_2}=\delta_{c_1c_2}$. To evaluate the
integral (\ref{IZHC}) for the case
of a matrix $\Gamma$ of reduced rank $M<N$ (i.e.\ when only $M$  eigenvalues denoted by $\gamma_1, \ldots , \gamma_M$ are nonzero, the rest $N-M$ being zero) we notice that the combination $\mbox{Tr} \,{K} \,{U}^{\dagger}{\Gamma} {U}=\sum_{c=1}^M \gamma_c\,{\bf u}^{\dagger}_c\hat{K}{\bf u}_c$ obviously depends only on the vectors ${\bf u}_c$ with $1\le c\le M$. These column vectors form a $N\times M$ matrix which we denote $\hat{V}_M$, and the ($M-$ component) rows of this matrix we denote as ${\bf v}^{\dagger}_j\,,\, j=1,...,N$. Then
$\mbox{Tr} {K} {U}^{\dagger}{\Gamma} {U}=\sum_{j=1}^Nk_j \left({\bf v}^{\dagger}_j
\,\hat{\gamma}_M{\bf v}_j\right)$ where we denoted $\hat{\gamma}=\mbox{diag}(\gamma_1,\ldots,\gamma_M)$.

Consider the $M\times M$ matrix $\hat{I}_M({\bf v})=\sum_{j=1}^N{\bf v}_j
\otimes {\bf v}^{\dagger}_j$. It is easy to see that the conditions of mutual
orthonormality of the $M$ vectors ${\bf u}_c\,,\,c=1,...,M$ are equivalent
to the condition of the matrix $\hat{I}_M({\bf v})$ to be just the identity matrix: $\hat{I}_M({\bf v})={\bf 1}_M$. Therefore the integration measure
over the manifold of matrices  $\hat{V}_M$ must be given by
\begin{eqnarray}\label{measureV}
\fl d\left[\hat{V}_M\right] \propto \prod_{j=1}^N
d^2{\bf v}_j \,\delta\left(\sum_{j=1}^N{\bf v}_j
\otimes {\bf v}^{\dagger}_j-{\bf 1}_M\right)
\\ \lo\nonumber
\propto \prod_{j=1}^N
d^2{\bf v}_j \int \left[d\hat{F}_M\right] e^{i\mbox{\small Tr}\hat{F}_M\left(\sum_{j=1}^N{\bf v}_j
\otimes{\bf v}^{\dagger}_j-{\bf 1}_M\right)}
\end{eqnarray}
where we used the (matrix) Fourier-representation for the (matrix) $\delta-$ function, with ${\hat F}_M$ being the manifold of Hermitian $M\times M$ matrices.

These facts allow us to rewrite the IZHC integral for the present case as:
\begin{eqnarray}
\fl \int\!{[dU]_N}\ e^{-i\,\mbox{\small Tr}{K}\hat{U}^{\dagger}{\Gamma}\hat{U}}\propto \int d\hat{F}_M e^{-i\mbox{\small Tr}\hat{F}}
\prod_{j=1}^N\int\left[d^2{\bf v}\right]
e^{i{\bf v}^{\dagger}\left(\hat{F}_M-k_j\hat{\gamma}_M\right){\bf v}}
\\ \nonumber \lo \propto
\int d\hat{F}_M e^{-i\mbox{\small Tr}\hat{F}_M}
\prod_{j=1}^N\det{\left(\hat{F}_M-k_j\,\hat{\gamma}_M\right)}^{-1}\ .
\end{eqnarray}
These   integrals to be well-defined we assume a positive infinitesimal positive imaginary part added to $\hat{F}_M$ in the denominator.
Now we introduce the Hermitian matrix $\hat{F}$ by
$\hat{F}_M=\hat{\gamma}_M^{1/2}\hat{F}\hat{\gamma}_M^{1/2}$ which yields the Jacobian factor:
$d\hat{F}_M=\left[\det{\hat{\gamma}_M}\right]^M\, d\hat{F}$. Diagonalizing $\hat{F}=U_M^{-1}\hat{\Lambda}_M U_M$ we use the
matrix of
eigenvalues $\hat{\Lambda}_M=\mbox{diag}{(\lambda_1,...,\lambda_M)}$ and
$M\times M$ unitary matrices $U_M$ as new variables of integration,
with the usual change in the measure as $d\hat{F}=\Delta^2(\hat \Lambda_M)d\hat \Lambda_M[dU]_M$.
After these manipulations we arrive at the relation
\begin{eqnarray}
\fl \nonumber\int\!{[dU]_N}\ e^{-i\, \mbox{Tr} {K} {U}^{\dagger}{\Gamma} {U}}\propto
\left[\det{\hat \gamma_M}\right]^{M-N}\int d\hat{\Lambda}_M
\Delta^2(\hat{\Lambda}_M)\prod_{j=1}^N\prod_{c=1}^M
\frac{1}{\lambda_c-k_j}\int\!{[dU]_M}\
e^{-i\, \mbox{Tr} \hat{\gamma}{U}_M^{\dagger}\hat{\Lambda}_M {U}_M}\\
\lo=\frac{\left[\det{\hat \gamma_M}\right]^{M-N}}{\Delta(\hat{\gamma}_M)}
\int d\hat \Lambda_M\Delta(\hat \Lambda_M)\prod_{c=1}^M\prod_{j=1}^N\frac{1}{\lambda_c-k_j}
\det{\left[e^{-i\gamma_{c_1}\lambda_{c_2}}\right]_{c_1,c_2=1, \ldots, M}}
\end{eqnarray}
where we used the standard IZHC formula Eq.(\ref{IZHC}) for the integration
over $U_M$. Finally, at the last step we use the symmetry of the
integrand with respect to  permutations of the arguments $\lambda_1,...,\lambda_M$
and arrive at the required extension of the IZHC integral (\ref{IZHCD}):
\begin{equation}\label{IZHCA}
\fl \int\,{[dU]_N}\ e^{-i\,\mbox{\small Tr}{K}{U}^{\dagger}{\Gamma}{U}}
\propto
M!\frac{\det{\gamma}^{M-N}}{\Delta({\gamma})}
\int d\hat{\Lambda}_M \Delta(\Lambda_M)\prod_{c=1}^M f_c(\lambda_c),
\end{equation}
where
\[
 f_c(\lambda_c)=\frac{e^{-i\gamma_c\lambda_c}}
{\prod_{j=1}^N\left(\lambda_c-k_j\right)}
\]
which is valid for the case when one of the matrices is of a reduced rank $M<N$.
Let us finally satisfy oneself that when both matrices $K$ and $\Gamma$ are of the full rank: $M=N$
our expression Eq.(\ref{IZHCA}) reproduces exactly the familiar Itzykson-Zuber formula (\ref{IZHC}). To this end we use the Cauchy identity:
$$
\det{\left(\frac{1}{\lambda_c-k_j}\right)}_{c,j=1,N}\equiv
\frac{\Delta(\Lambda)\Delta(K)}{\prod_{j=1}^N\prod_{c=1}^N
(\lambda_c-k_j)}
$$
which allows us to write the write-hand side of (\ref{IZHCA}) as
\begin{equation}\begin{array}{l}
\frac{i^{-N}}{\Delta(\Gamma)\Delta(K)}\int d\Lambda
e^{-\sum_{i=1}^N\gamma_i\lambda_i}
\det{\left(\frac{1}{\lambda_c-k_j}\right)}_{c,j=1,N}\\
 \equiv \frac{i^{-N}}{\Delta(\Gamma)\Delta(K)}
\sum_{T}(-1)^T\prod_{c=1}^N\int d\lambda_ce^{-i\gamma_c\lambda_c}\frac{1}{\lambda_c-k_{j_c}}\end{array}
\end{equation}
where the summation goes over all permutations $T=\left(k_{j_1},...,k_{j_N}\right)$ of the set $(k_1,...,k_N)$,
with the factors $(-1)^T$ standing for the signs of permutation.
Each integration yields a factor: $i\exp({-i\gamma_ck_{j_c}})$.
Taking into account the factors $(-1)^T$ allows one to present the
result as a conventional Itzykson-Zuber determinant.

\subsection{Integration over unitary group for subunitary matrices}
\label{sec:A}
We start from Eq.(\ref{P({z})q}) and want to perform all integrations over matrices ${\hat Q}_l$ and complex vectors ${\bf a}$.
Rescaling all variables according to $\tilde Q_i = \sqrt {\hat T}\  \hat Q_i\  \sqrt {\hat T}$ with $\tilde Q_0 = {\bf 0}$ and  $\tilde Q_N = \hat T$, and $\tilde {\bf a} =  \sqrt {\hat T}\  {\bf a}$  and taking into account all Jacobians from the transformations we arrive at
\begin{eqnarray}\label{P({z})Q} \nonumber
{\cal P}(\{Z\}) &\propto&  |\Delta(\{Z\})|^2
\int \prod_{i=1}^{N-1} d\tilde Q_i\   \prod_{i=1}^N \int d^2\tilde {\bf  a} \  \delta(\tilde Q_i-\tilde Q_{i-1}-\tilde {\bf a} \otimes {\tilde {\bf a}}^{\dagger})\\
  &\times& \prod_{l=1}^N
\delta\left(|z_1|^2...|z_l|^2-
\det\left[{\bf 1}-\tilde Q_l \right]\right)\ \det\hat T^{M-N}\ .
\end{eqnarray}
The following strategy is to diagonalize all matrices $\tilde Q_i$ as $\tilde Q_i = U_i \lambda_i {U_i}^{-1}$ with  $$d\tilde Q_i \propto \prod_{j<k} (\lambda_i^{(j)} - \lambda_i^{(k)})^2\ \prod_l d\lambda_i^{(l)} \ d\mu(U_i) $$
and then perform the integrations over the eigenvectors $U_i$.
We start with $\int d^2{\bf a} \ \delta( \tilde Q_1 - {\bf a} \otimes {\bf a ^{\dagger}}) $. Since we integrate over all directions of the complex vector ${\bf a}$ the integral does not depend on $U_1$, i.e. the eigenvectors of $\tilde Q_1$ and is given by (cf. Eq.(\ref{measureV}):
\begin{eqnarray}  \nonumber
\fl \int d^2{\bf a}\ \delta( \tilde Q_1 - {\bf a} \otimes {\bf a ^{\dagger}})
 \propto  \int d\Omega \exp(i {\rm Tr} \Omega \tilde Q_1)/ \det(\epsilon +i\Omega)\\ \label{Q_1}
\propto {1\over \prod_{j<k} (\lambda_1^{(j)} - \lambda_1^{(k)})} \prod_{j<k}\left(-{\partial \over \partial \lambda_1^{(j)}} + {\partial \over \partial \lambda_1^{(k)}}\right)\ \prod_l \theta(\lambda_1^{(l)})\ .
\end{eqnarray}
 Here we diagonalized the matrix $\Omega$  and integrated out the corresponding
eigenvectors with the use the IZHC formula Eq.(\ref{IZHC}), with
$\epsilon >0$ ($\propto$ unity matrix) being an infinitesimal regularisation matrix. The result  is a complicated expression in terms of $\delta$-functions and step functions. We now proceed to calculate the next factor in (\ref{P({z})Q}). Using again the IZHC integral two times we obtain
\begin{equation} \label{Q_2}
\fl \int d\mu (U_1)\int d^2{\bf a}\ \delta( \tilde Q_2 - \tilde Q_1-{\bf a} \otimes {\bf a ^{\dagger}})  \propto  {\det ( \theta ( \lambda_2^{(l)} - \lambda_1^{(m)})) \over \prod_{j<k} (\lambda_2^{(j)} - \lambda_2^{(k)})(\lambda_1^{(j)} - \lambda_1^{(k)})}
\end{equation}
We used here that (\ref{Q_1}) was independent of $U_1$ and we see that (\ref{Q_2}) is independent of $U_2$. Using the latter fact on the next stage, we are able in this way to perform the integrations over all $\tilde Q_i$-diagonalizing matrices $U_i$. There remain the integrations over all the eigenvalues $\lambda_i^{(j)}$ of the matrices $\tilde Q_i$.
The resulting expression for ${\cal P}(\{Z\})$ may now be written as
\begin{eqnarray}\label{P({z})Q_i} \nonumber
\fl {\cal P}(\{Z\}) \propto {\det\hat T^{M-N} \over \prod_{j<k}(T_j-T_k)} |\Delta(\{Z\})|^2
\int \prod_{i=1}^{N-1} dQ_i\ \theta (\hat T- Q_{N-1})...\theta(Q_2-Q_1) [\Delta_1 \theta(Q_1)]  \\
  \times \prod_{l=1}^N
\delta\left(|z_1|^2...|z_l|^2-
\det\left[{\bf 1}- Q_l \right]\right)\   .
\end{eqnarray}
with diagonal $M \times M$ matrices $Q_i$, and the notations
$dQ_1 := \prod_{i=1}^M d\lambda_1^{(i)}$
\begin{eqnarray}
\fl \theta ( Q_2 -Q_1 ):=\prod_{i=1}^M \theta ( \lambda_2^{(i)}  -\lambda_1^{(i)} )\quad,\quad \det({\bf 1}-Q_1):=\prod_{i=1}^M (1-\lambda_1^{(i)})
\\ \lo  \Delta_1 \theta ( Q_1) :=\prod_{j<k}^{1...M}\left(-{\partial \over \partial \lambda_1^{(j)}} + {\partial \over \partial \lambda_1^{(k)}}\right)\ \prod_{l=1}^M \theta(\lambda_1^{(l)})\ .
\end{eqnarray}
An important simplification occurs if we go to logarithmic variables $\tau_i = \ln({\bf 1}-Q_i)$ (again these are diagonal $M \times  M$ matrices with $0\ge \tau_i \ge \ln({\bf 1}-\hat T) = \tau_N$ ). We have
\begin{eqnarray}  \nonumber
\fl \delta\left(|z_1|^2...|z_l|^2-
\det\left[{\bf 1} - Q_l \right]\right) \theta (Q_l - Q_{l-1}) = {1\over \det({\bf 1}-Q_l)}\ \delta\left(\sum_{j=1}^l \ln |z_j|^2- \sum_{k=1}^M \ln(1-\lambda_l^{(k)})
 \right)\\ \nonumber
\times \prod_{j=1}^M \theta( \ln(1-\lambda_{l-1}^{(j)}) - \ln (1- \lambda_{l}^{(j)}))
\end{eqnarray}
and there is the special relation
$$  \Delta_1 \theta (Q_1) = \prod_{j<k}^{1...M}\left({\partial \over \partial \tau_1^{(j)}} - {\partial \over \partial \tau_1^{(k)}}\right)\ \prod_{l=1}^M \theta( - \tau_1^{(l)}) =: \tilde \Delta_1 \theta(-\tau_1) \ . $$
Taking into account the Jacobian from the transformation our joint density
takes the form
\begin{eqnarray}\label{P({z})tauN} \nonumber
\fl {\cal P}(\{Z\}) \propto {\det {\hat T}^{M-N} |\Delta(\{Z\})|^2 \over \det({\bf 1}-\hat T ) \prod_{j<k}(T_j-T_k)}
\int \prod_{i=1}^{N-1} d\tau_i\ [\tilde \Delta_1 \theta(-\tau_1)]\  \theta (\tau_1- \tau_2)...\theta (\tau_{N-1}- \tau_N)      \\
  \times \prod_{l=1}^N
\delta\left(\ln |z_1|^2+...+\ln |z_l|^2-
{\rm Tr}\tau_l \right)\   .
\end{eqnarray}
Now we see that the operator $\tilde \Delta_1$ can be shifted by partial integration, it commutes with ${\rm Tr}\tau_1$ and then acts as $\tilde \Delta_2$ on $\theta (\tau_1- \tau_2)$. This argument can be repeated, such that finally the operator $\tilde \Delta_N$ appears in front of the remaining integral. Changing again the integration variables we obtain
\begin{equation}\label{P({z})tau}
\fl {\cal P}(\{Z\})\propto {\det {\hat T}^{M-N} |\Delta(\{Z\})|^2 \over \det({\bf 1}-\hat T ) \prod_{j<k}(T_j-T_k)}
\tilde \Delta \int \prod_{l=1}^{N}\left [d\tau_l\   \theta(-\tau_l)\delta\left( \ln |z_l|^2- {\rm Tr}\tau_l \right)\right ]
    \delta(\sum_{i=1}^N\tau_i  -\tau )\
\end{equation}
with $\tau := \ln({\bf 1}-\hat T)$ and
$\tilde \Delta := \prod_{j<k}^{1...M}\left({\partial \over \partial \tau^{(j)}} - {\partial \over \partial \tau^{(k)}}\right)\ .$
Writing down the Fourier representations of the $\delta$-functions in (\ref{P({z})tau})  the integrations over the diagonal $M \times M$ matrices   $\tau_i$ can mainly be performed yielding
\begin{equation}\label{omega} \nonumber
\fl \int_{\{\tau_i <0\}} \prod_{l=1}^{N}\left [ d\tau_l\   \delta\left( \ln |z_l|^2- {\rm Tr}\tau_l \right) \right ] \delta\left(\sum_{i=1}^N\tau_i -\tau \right)\ =
\int\left[ {d\hat \omega \over 2\pi}\right]\exp(-i{\rm Tr}\hat \omega\ \tau ) \prod_{l=1}^Nf(\ln|z_l|^2, \hat \omega)
\end{equation}
with
\begin{eqnarray}\label{f(a,omega)}
\fl f(a,\hat \omega )= \int { d\Omega \over 2\pi} {\exp(i\Omega a) \over \prod_{k=1}^M (-i\Omega + i\omega_k + \epsilon )}
  = \sum_{k=1}^M{\exp(i\omega_k a) \over \prod_{j (\ne k)} (-i\omega_k + i\omega_j ) } \theta(-a)\
\end{eqnarray}
and with an infinitesimal $\epsilon >0$. The integration over $\Omega$ can be performed in the complex plane yielding the restriction $a<0$, i.e. in (\ref{P({z})tau}) all $z_i$ lying inside the unit circle ($\ln|z_i|^2<0$).
Thus we have derived the full joint density of eigenvalues of subunitary matrices (\ref{main}), which can be used to derive all correlation functions. 

\section{Derivation of Eq.(\ref{interm})}
Notice that in the
limit $N\to\infty$ we can replace $N$ in the exponent in Eq.\
(\ref{P(J1)}) by $N-n$. Then, similar to the Gaussian case,
see Eq.(\ref{example}), we first write:
\begin{eqnarray}\label{FF}
\fl R_n(\{z\}) \propto \frac{e^{-\frac{N-n}{2}\left[\sum_{j=1}^n \re\,
z_j^2+ \sum_{c=1}^M \gamma_c^2\right]}|\Delta(\{z\})|^2}
{\Delta({\gamma}) \det {\gamma}^{N-M}} 
\!\!\int\!\!d^2\{\xi\}\
e^{-\frac{N-n}{2}\!\! \sum\limits_{l=1}^{N-n}\!\! \re\, \xi_l^2
}|\Delta (\{\xi\})|^2
\!\!\!\\
\lo \nonumber
\times
\prod_{l=1}^{N-n}\prod_{j=1}^n|z_j-\xi_l|^2
\int\! d\Lambda\
 a(\Lambda,{\gamma})
\Delta (\Lambda)\prod\limits_{l=1}^{N-n}
\sum\limits_{c=1}^M\frac{e^{i\im \xi_l\lambda_c}}
{\prod\limits_{s(\ne c)}(\lambda_c-\lambda_s)}
\end{eqnarray}
where
\begin{eqnarray}\label{A}
\fl a(\Lambda,{\gamma})=e^{-i\sum_{c=1}^M\tilde{\gamma}_c\lambda_c}
\prod\limits_{j=1}^n\sum\limits_{\kappa=1}^M
\frac{e^{i\im z_j\lambda_{\kappa}}}
{\prod\limits_{s(\ne \kappa )}(\lambda_{\kappa}-\lambda_s)}=
\sum_{\kappa_1=1}^M\ldots\!\sum_{\kappa_n=1}^M
A_{{\kappa}}(\Lambda,{\gamma})\\ \lo\nonumber
A_{{\kappa}}(\Lambda,{\gamma})
\!=\!\frac{e^{-i\sum_{c=1}^M\tilde{\gamma}_c\lambda_c}}
{\prod\limits_{j=1}^n\prod\limits_{s(\ne \kappa_j )}^M
(\lambda_{\kappa_j}-\lambda_{s})}\ .
\end{eqnarray}
In Eq.\ (\ref{A}) we denoted
${\kappa}=(\kappa_1,\ldots,\kappa_n)$ and $\tilde{\gamma}_c=
\gamma_c-\sum_{j=1}^n\delta_{c,\kappa_j}\im z_j$, where $\delta_{c,\kappa_j}=1$ if
$c=\kappa_j$  and zero otherwise. This allows us to write:
\begin{equation}\label{RRR}
R_n(\{z\}) \propto  \frac{e^{-\frac{N-n}{2}\sum_{j=1}^n \re\,
z_j^2}|\Delta(\{z\})|^2}{\det{\gamma}^n}
\sum_{\kappa_1=1}^M\ldots\!\sum_{\kappa_n=1}^M
\!{F}_{{\kappa} } ({\gamma}, \{z\}),
\end{equation}
where
\begin{eqnarray}\label{F}
\fl {F}_{\!{\kappa}}({\gamma}, \{z\})
\!=\! B({\gamma})\!\!\int\!\!d^2\{\xi\}\
e^{-\frac{N-n}{2}\!\! \sum\limits_{l=1}^{N-n}\!\! \re\, \xi_l^2
}|\Delta (\{\xi\})|^2
\!\!\!\prod_{l=1}^{N-n}\prod_{j=1}^n|z_j-\xi_l|^2\\ \nonumber
\times  \int\! d\Lambda\
 A_{{\kappa}}(\Lambda,{\gamma})
\Delta (\Lambda)\prod\limits_{l=1}^{N-n}
\sum\limits_{c=1}^M\frac{e^{i\im \xi_l\lambda_c}}
{\prod\limits_{s(\ne c )}(\lambda_c-\lambda_s)}
\end{eqnarray}
and we denoted
\[
B({\gamma})\!  =\! \frac{e^{-\frac{N-n}{2}\sum_{c=1}^M  \gamma_c^2}}
{\Delta({\gamma}) \det{\gamma}^{N-n-M}} \!\!\ .
\]

Now we introduce the differential operators
$\hat{D}_{\kappa}=\prod_{j=1}^{n}\prod_{s(\ne \kappa_j )}^M \left(
\frac{\partial}{\partial \gamma_{k_j}}-\frac{\partial}{\partial
\gamma_s}\right)$ and consider its action on the ratio
$F_{{\kappa}}({\gamma})/B({\gamma})$ from Eq.(\ref{F}).
On one hand, we notice that:
$$
\hat{D}_{\kappa} A_{{\kappa}}({\lambda},{\gamma})=
(-i)^{n(M-1)} e^{-i\sum_{c=1}^M\tilde{\gamma}_c\lambda_c}.
$$ 
and, as a result the action of $\hat{D}_{\kappa}$ on
$F_{{\kappa}}({\gamma})/B({\gamma})$ brings the ${\sf \xi}$-integrand
exactly to the form that allows us to rewrite it as
an average of a product of determinants
(cf.Eq.(\ref{example})). The averaging goes over the density of joint
probability of eigenvalues $\xi_1,\xi_2,...,\xi_{N-n}$ of a
non-Hermitian matrix $J_{N-n}(\tilde{{\gamma}})=
H_{N-n}+i\tilde{\Gamma}$, with nonzero eigenvalues of
$\tilde{\Gamma}$ being $\tilde{\gamma_1},...,\tilde{\gamma_M}$:
\begin{equation}\label{C}
 \fl \hat{D}_{\kappa}
\left[F_{{\kappa}}({\gamma})/B({\gamma})\right]=
 e^{\frac{N-n}{2} \sum_{c=1}^M  \tilde{\gamma}_c^2}
\Delta({\tilde{\gamma}}) \det\!\!{\tilde{\gamma}}^{N-M}
\left\langle\prod_{j=1}^n
\Big| \det
\left[z_j-{J}_{N-n}(\tilde{{\gamma}})\right]\Big|^2
\right\rangle_{GUE}
\end{equation}

On the other hand, to ensure existence of a well-defined
limit $N\to \infty$ of the correlation functions $R_n(\{z\})$ in
Eq.(\ref{RRR}) the quantities
$F_{{\kappa}}({{\gamma}})$ must in that limit behave regularly as
 functions of $\gamma_c$, i.e.
$F_{{\kappa}}({\gamma})|_{N\gg 1}\sim
C_N F^{(s)}_{{\kappa}}({\gamma})$ where $C_N$ does not
depend on ${\gamma}$ and  $F^{(s)}_{{\kappa}}({\gamma})$
 has a finite limit when
$N\to\infty$. Because of this property in the limit
$N\gg n$ the action
of $\hat{D}_{\kappa}$ on
$$F_{{\kappa}}({\gamma})/B({\gamma})
\equiv F_{{\kappa}}({\gamma})
\Delta({\gamma})\det{\gamma}^{N-n-M}
e^{\frac{N-n}{2} \sum_{c=1}^M  \gamma_c^2}$$
to the leading order amounts to:
\begin{equation}\label{D2}
\hat{D}_{\kappa}
\left[F_{{\kappa}}({\gamma})/B({\gamma})\right]\approx
F_{{\kappa}}({\gamma})\Delta({\gamma})
\hat{D}_{\kappa}\left[\det{\gamma}^{N-n-M}
e^{\frac{N-n}{2} \sum_{c=1}^M  \gamma_c^2}\right]\ .
\end{equation}
Indeed, each $\gamma$-differentiation of the terms in the square brackets
in Eq.(\ref{D2}) brings a factor of the order of $N$, which is much larger
compared to the result of differentiating the factor
$F_{{\kappa}}({\gamma}) \Delta({\gamma})$.
Performing the remaining differentiations explicitly and considering
$N-n\approx N-n-M\approx N$, we find:
\begin{eqnarray*}
\fl
\hat{D}_{\kappa}
\left[F_{{\kappa}}({\gamma})/B({\gamma})\right]|_{N\gg n,M}\approx
(2N)^{n(M-1)} F_{{\kappa}}({\gamma})
\Delta({\gamma})\\ \nonumber \times
\det{\gamma}^{N-n-M}
e^{\frac{N-n}{2} \sum_{c=1}^M  \gamma_c^2}
\prod_{j=1}^n\prod_{s(\ne \kappa_j )}^M\left(g_{\kappa_j}-g_s\right)
\end{eqnarray*}
where we introduced the notation $g_c=\frac{1}{2}(\gamma_c+\gamma_c^{-1})$.

Comparing this expression with Eq.(\ref{C}) yields
\begin{equation}\label{nn}
\fl {F}_{{\kappa}}({\gamma}, \{z\})\propto
\frac{{C}_{\tilde{{\gamma}}}(\{
z\})}{\prod_{j=1}^{n}\prod_{s(\ne \kappa_j )}^M (g_{\kappa_j}-g_{s})}
 \frac{\Delta (\tilde{{\gamma}})} {\Delta
({\gamma})} \!\! \left[\det
\frac{\tilde{{\gamma}}}{ {\gamma}}\right]^{N-n-M}
 e^{-\frac{N-n}{2}
\sum_{c=1}^M\left(\gamma_c^2-\tilde{\gamma}_c^2\right)}
\end{equation}
where we denoted:
\begin{equation}\label{CC}
{C}_{\tilde{{\gamma}}}(\{z\})=
 \left\langle\prod_{j=1}^n
\Big| \det
\left[z_j-{J}_{N-n}(\tilde{{\gamma}})\right]\Big|^2
\right\rangle_{GUE}\ .
\end{equation}

In the limit when $N\to\infty$ and $M$  finite, almost all
eigenvalues of $J$ have imaginary part of the order $\frac{1}{N}
\ll \gamma_c$ \cite{note1}. Rescaling the imaginary parts $y_j=N\im
z_j$ one finds that
\[
\frac{\Delta (\tilde{{\gamma}})} {\Delta ({\gamma})}
\left[\det \frac{\tilde{{\gamma}}}{
{\gamma}}\right]^{N-n-M}\!\! \!\!\!e^{-\frac{N-n}{2}
\sum_{c=1}^M\left(\gamma_c^2-\tilde{\gamma}_c^2\right)}=
e^{-2\sum_{j=1}^n y_jg_{\kappa_j}}
\]
in the limit $n,M\ll N\to \infty$. As will be clear later on,
 in the same limit one can
substitute ${\gamma}$ for $\tilde{{\gamma}}$ in the GUE
averaged product of determinants. Combining all the factors we arrive at Eq.(\ref{interm}).\\

\end{document}